# Bloated Stars as AGN Broad Line Clouds: The Emission Line Spectrum

Tal Alexander and Hagai Netzer
*School of Physics and Astronomy and the Wise observatory*⋆, *Tel-Aviv University, Tel-Aviv 69978, Israel.*

## ABSTRACT

The 'Bloated Stars Scenario' proposes that AGN broad line emission originates in the winds or envelopes of bloated stars (BS). Its main advantage over BLR cloud models is the gravitational confinement of the gas and its major difficulty the large estimated number of BSs and the resulting high collisional and evolutionary mass loss rates. Previous work on this model did not include full calculations of the spectrum and used very simplified stellar distribution functions. Here we calculate the emission line spectrum by applying a detailed numerical photoionization code to the wind and by assuming a detailed nucleus model (Murphy, Cohn & Durisen 1991). We study a wide range of wind structures for a QSO model with $L_{\rm ion} = 7{\times}10^{45}\,$erg/s, $M_{\rm bh} = 8{\times}10^7 M_\odot$ and a stellar core density of $7{\times}10^7 M_\odot/\,{\rm pc}^3$. The BSs emit a spherically symmetric wind whose size and boundary density are determined by various processes: Comptonization by the central continuum source, calculated here self consistently, tidal disruption by the black hole and the limit set by the wind's finite mass. We find that the emission spectrum is mainly determined by the conditions at the boundary of the line emitting fraction of the wind rather than by its internal structure. Comptonization results in a very high ionization parameter at the boundary which produces an excess of unobserved broad high excitation forbidden lines. The finite mass constraint limits the wind size, increases the boundary density and thus improves the results. Slow, decelerating, mass-constrained flows with high gas densities ($10^8$ to $10^{12}$ cm$^{-3}$) are as successful as cloud models in reproducing the overall observed line spectrum. The Mg II $\lambda$2798 and N V $\lambda$1240 lines are however under-produced in our models. We adjust the number of BSs so as to obtain the observed $EW({\rm L}\alpha)$ and find that only $\sim 5\times10^4$ BSs with dense winds are required in the inner 1/3 pc. This small fraction approaches that of supergiants (SG) in the solar neighborhood and the calculated mass loss is consistent with the observational constraints. The required number of BSs, and their mass loss rate, are very sensitive functions of the wind density structure. SG-like BSs are ruled out by their high mass loss. BSs with dense winds can reproduce the BLR line spectrum and be supported by the stellar population without excessive mass loss and collisional destruction rates. The question whether such hitherto unobserved stars actually exist in the BLR remains open.

## 1 INTRODUCTION

The observed properties of active galactic nuclei (AGN) lead to the conclusion that the line emission originates in numerous small, cold and dense gas concentrations, denoted by the general term *clouds.* It is inferred that the Broad Line Region (BLR) cloud system typically consists of $\sim 10^7$ clouds with particle densities of $10^8 - 10^{12}$ cm$^{-3}$ and an ionized fraction of dimension $3{\times}10^{12}$ cm at a temperature of $\sim 10^4\,$K. However, the clouds' distribution and velocity field, as well as the functional dependence of a cloud's density and size on the distance from the black hole, cannot be deduced from present day observations and are therefore free parameters of the cloud model. Once these parameters are specified by additional assumptions, the BLR line emission can be calculated by photoionization codes. These calculations demonstrate that such cloud models can reproduce to a fair degree the observed line properties (Netzer 1990).

⋆ Of the Beverly and Raymond Sackler Faculty of Exact Sciences.

The major objection to the clouds model is the confinement problem. The clouds, with estimated masses of $10^{-9} M_\odot$, are not self gravitating objects and should disintegrate on the sound crossing time scale of $\sim 0.5$ yrs, whereas the crossing time of the BLR is $\sim 30 L_{46}^{1/2}$ yrs (where $L_{46}$ is the ionizing continuum luminosity in units of $10^{46}$ erg/s). Unless clouds are continuously formed in the BLR, an external confinement mechanism is necessary. A once promising idea, the two-phase model (Krolik, McKee & Tarter 1981) whereby pressure equilibrium is maintained between the cloud (cold phase at $\sim 10^4$ K) and the hot interstellar medium (HIM, hot phase at $\sim 10^8$ K) has met serious objections. The gas temperature in such models depends on the shape of the UV–X-ray continuum. Realistic continua models predict the HIM temperature to be $\sim 10^7$ K which is not compatible with two-phase equilibrium (Fabian et al 1986; Mathews & Ferland 1987). Additional problems are cloud breakup by drag forces due to the supersonic motion of the clouds, the unobserved opacity expected from this cooler HIM and the instability of the cold phase due to internal line pressure.

An alternative model that has not been investigated in detail yet is the 'stellar winds scenario' (Edwards 1980; Mathews 1983; Scoville & Norman 1988; Penston 1988; Kazanas 1989). This scenario proposes that the lines are emitted from the mass-loss winds or envelopes of supergiants. and is motivated by the following arguments: 1) The confinement problem is resolved by introducing a self gravitating gas reservoir—the star—that can continuously replenish the cold gas that is evaporating to the HIM. 2) Stars are expected to exist near the central source and estimates suggest that the conditions in supergiant winds resemble those inferred for the line emitting gas. 3) The lack of evidence for radial velocity of the BLR gas (e.g. Maoz et al. 1991) is consistent with the isotropic velocity field of the virialized stellar core around a massive black hole.

Earlier work on stellar wind models was confined to simplified approximations. In particular, there are yet no detailed calculations of the line ratios and the line profiles were approximated using unrealistic stellar distributions and line emissivity. The most serious objection to the model as yet is the large number of supergiants needed to produce the line luminosity. The resulting high stellar density leads to an unacceptably large collisional destruction rate and to a very high mass injection rate into the HIM which is precluded by observations (Begelman & Sikura 1991). These unfavorable estimates depend strongly on assumptions about the stellar distribution in the galactic core, estimates of the BLR size and the size and density structure of the winds.

In this paper we will concentrate on the questions of the line ratios, the number of bloated stars required for obtaining the covering factor and the size of the BLR. The issues relating to the dynamics of the BLR and the line profiles will be discussed elsewhere. In § 2 we specify the properties of the stellar core and of the winds of the bloated stars and discuss the assumptions and approximations underlying our models. In § 3 we describe the photoionization code, its application to the winds and the calculations of the emission line spectrum. In § 4 we discuss the various observational and theoretical constraints on the wind models. In § 5 we present results for a single bloated star and for the entire bloated star population. In § 6 we discuss the implications for the bloated stars model and the constraints that can be placed on it.

## 2 THE STELLAR CORE AND BLOATED STAR MODEL

Our aim is to model the stellar winds scenario and to calculate the expected emission line spectrum. Two ingredients are necessary: the stellar distribution function, $n_\star$, and the properties of the super giant winds. In modeling the stellar distribution function we will use the results of detailed numeric calculations that follow the evolution of a stellar core around an initial black hole and the resultant black hole growth and continuum radiation evolution (Murphy, Cohn & Durisen 1991, hereafter MCD). Since giant stars in AGN core may be very different from those observed in the solar neighborhood, we adopt the general term *bloated stars* rather than the more specific term supergiants for designating the wind emitting objects. The bloated stars will be modeled by simplified wind structures with no claim to hydrodynamical self-consistency. The bloated stars model differs in several significant aspects from the clouds model: The wind gas density is not uniform but changes with the outflowing gas, The distribution of bloated stars is expected to be correlated to the density of the galactic core (and possibly also to the irradiating flux) and the velocity field is that of the stellar core.

### 2.1 The stellar core

The attempts of previous works to use a dynamically self consistent stellar core distribution were based on a stellar distribution of the form $n_\star \propto r^{-7/4}$ (e.g. Kazanas 1989). This distribution can be derived from simple dimensional arguments of energy conservation in the equilibrium state of a collisionless system of point masses in a deep potential well, isotropic both in position and velocity space (see e.g. Binney & Tremaine 1987). The immediate stellar surrounding of AGN is probably isotropic and virialized irrespective of the host galaxy type. However, the omission of star–star collisions is not justified in the dense environment of the deep potential well.

In this work we use the results of a much more detailed and realistic analysis performed by MCD. Their work follows the dynamical and luminosity evolution of a galactic nucleus consisting of a dense coeval stellar system surrounding a massive black hole. The stellar population contains stars of masses from 0.3 to $30 M_\odot$ which are initially distributed according to a power-law initial mass function (IMF):

$$dn_\star(m) \propto m^{-(1+x)} dm \, , \qquad (1)$$

where $dn_\star$ is the number of stars in the mass interval $dm$. The black hole is fueled by mass loss due to tidal disruption of stars in the loss cone, inelastic (physical) collisions of stars, and stellar evolution. The mass loss due to collisions is treated in detail taking into account the radii, masses, relative velocity and impact parameter of the colliding stars. However, collisions with evolved stars are not taken into account. Mergers and the formation of binaries are also not considered. It is assumed that all the mass released by these processes accumulates in a gas reservoir near the black hole and is eventually completely accreted. It is implicitly assumed that the accretion is spherical and limited by the Eddington accretion rate. The initial conditions are a seed black hole of $M_{\rm bh} \sim 10^4 M_\odot$, a Plummer stellar distribution function with initial velocity equipartition and the power-law index of the IMF. The isotropy in coordinate and velocity space is maintained throughout the calculations.

The MCD results show that in low density cores ($\lesssim 7\times10^6\ M_\odot/\,{\rm pc}^3$), where elastic collisions dominate, the core expands due to energy released by stars captured in tightly bound orbits and a $n_\star \propto r^{-7/4}$ density power-law is established, as discussed above. In higher density cores ($\gtrsim 7\times10^7\ M_\odot/\,{\rm pc}^3$), the inelastic collisions lead to core contraction and a $n_\star \propto r^{-1/2}$ density power-law. In all models the central region is eventually depleted by stellar disruptions and collisions. This is more pronounced in high density cores due to the larger black hole mass and the higher frequency of collisions. At large distances ($> 1$ pc), where the stellar dynamic time-scale is very long, the density function essentially reflects the initial Plummer law stellar density and is therefore to some extent arbitrary. MCD also show that the density functions of the different mass groups, all initially the same, evolve differently ('mass segregation').

The calculations reproduce the full range of observed AGN luminosities from $10^{43}$ to $10^{48}$ erg/s by varying the initial core density and IMF power-law index. The luminosity is calculated on the assumption of a constant mass conversion efficiency of 0.1. Initially limited by the Eddington luminosity, it grows exponentially, reaches a sharply defined peak luminosity after a $few\times10^8$ yrs and then declines slowly. The black hole growth is likewise limited by the Eddington limit until the peak luminosity is reached at which time the accretion rate equals the replenishment rate of the gas reservoir. From that point on the gas reservoir is rapidly exhausted and the accretion of inflowing gas proceeds immediately at a rate governed by the mass loss processes in the core (see figures 5 & 10 in MCD). The peak luminosity and the final mass of the black hole increase with the initial core densities.

A full study of the bloated stars scenario should in principle check separately each of the core configurations described above. Our work, being an initial study, will use one particular core model in the attempt to reproduce the line emission of a typical AGN. The choice of this core model, as discussed below, is an attempt to model an AGN which is both representative of its class and has the optimal properties for the bloated stars scenario as suggested by the previous works on this subject.

Previous estimates (Scoville & Norman 1988; Kazanas 1989; Begelman & Sikura 1991) predict that the number of BSs in the core may not be enough for the required BLR covering factor of about 0.1. It is therefore preferable to choose a core model that has the largest number of stars for a given continuum luminosity. The MCD results indicate that the more massive cores have a larger fraction of their mass locked in the black hole and consequently fewer stars as compared to less massive cores. On the other hand less massive cores may not be luminous enough for bright AGNs. Our choice is to model an intermediate mass core that reaches bright Seyfert 1 or low quasar luminosities but still has a significant fraction of its mass in the stars.

It is necessary to specify the core's evolutionary stage since the shape of the stellar distribution changes with time. The initial stage of exponential luminosity growth is too short-lived ($few\times10^7$ yrs) and not luminous enough. The choice between the two subsequent stages, peak luminosity and post-peak power-law decline, is less obvious. The peak luminosity lasts a $few\times10^8$ yrs at which time the stellar density is relatively peaked at the center. This may improve the covering factor problem and tend to decrease the BLR size (in step with recent reverberation studies) but, on the other hand, may also increase the collision rate. This choice is problematic since it implies that the lifetime of a single AGN is short and consequently there should be many galaxies with relic supermassive black holes at their nucleus to account for the observed space density of AGN (Blandford 1990).

The alternative of modeling the core at the post-peak stage has the advantage that it is the stage where the AGN spends most of its lifetime and that the core's structure and luminosity change relatively slowly so that the modeling is not strongly time dependent†. During the post-peak stage the core is depleted, the collision rate is relatively low and the covering factor is expected to be smaller.

In this work we chose to model a core just before its peak luminosity. This model (MCD model 2B) has an IMF index $x = 1.5$ (eq. 1) and begins with a central density of $7.2\times10^7 M_\odot/\,{\rm pc}^3$, a core mass of $3\times10^8 M_\odot$ within the inner 1 pc and a total nucleus mass of $8.5\times10^8 M_\odot$ within 100 pc. The luminosity peaks at $4.0\times10^8$ yrs with $L_{peak} = 2.4\times10^{46}$ erg/s and a black hole mass of $M_{\rm bh} = 1.9\times10^8 M_\odot$. We model the cluster at $3\times10^8$ yrs at which time the accretion rate is at the Eddington limit with a bolometric luminosity of $10^{46}$ erg/s, the black hole mass is $8\times10^7 M_\odot$ and the mass loss from the core to the central gas reservoir proceeds at a rate of $0.3 M_\odot/\,{\rm yr}$. The stellar distribution is taken to be that of the $0.8 M_\odot$ mass bin on the assumption that stars of about solar mass are representative of the BSs progenitors (§ 2.2) (the distribution functions of the various stellar mass bins vary somewhat in their radial scale but their overall shapes are quite similar).

The stellar density distribution exhibits the three distinct zones discussed above: depletion near the black hole, a $r^{-1/2}$ power-law at intermediate radii and steepening at large radii. A simplified form (Figure 1) was interpolated between stellar distributions given by MCD at $2.8\times10^7$ and $1.04\times10^9$ yrs and was used to approximate the stellar distribution. The interpolation was done linearly in $\log(t)$ and $\log(n_\star)$ since the evolution of $\sim 1 M_\odot$ stars at $\sim 10^8$ yrs is governed by stellar collisions and disruption by the black hole, both roughly linear in the $\log(t) - \log(\dot{M})$ plane. This distribution contains $7\times10^5 M_\odot$ in the inner 0.1pc, $8\times10^7 M_\odot$ in the inner 1 pc and a total of $2\times10^8 M_\odot$ which is $\sim 1/4$ of the total stellar mass of the system.

The continuum spectrum used in this work is a synthetic spectrum typical of AGNs (e.g. Rees, Netzer & Ferland 1989) with $L_{\rm ion} = 7\times10^{45}$ erg/s and $L_{bol} = 10^{46}$ erg/s and mean ionizing photon energy of $\langle h\nu \rangle = 2.58\,$Ryd. We do not specify the origin of this continuum and assume that all the continuum radiation is emitted *isotropically* from a point-like source. The possibilities of an anisotropic source or obscuration are not considered.

† MCD conclude that the slow luminosity decrease of their models is inconsistent with the observed redshift dependence of AGN luminosity and suggest a decreasing mass conversion efficiency as a way of obtaining a steeper luminosity decline. This is beyond the scope of the present work.

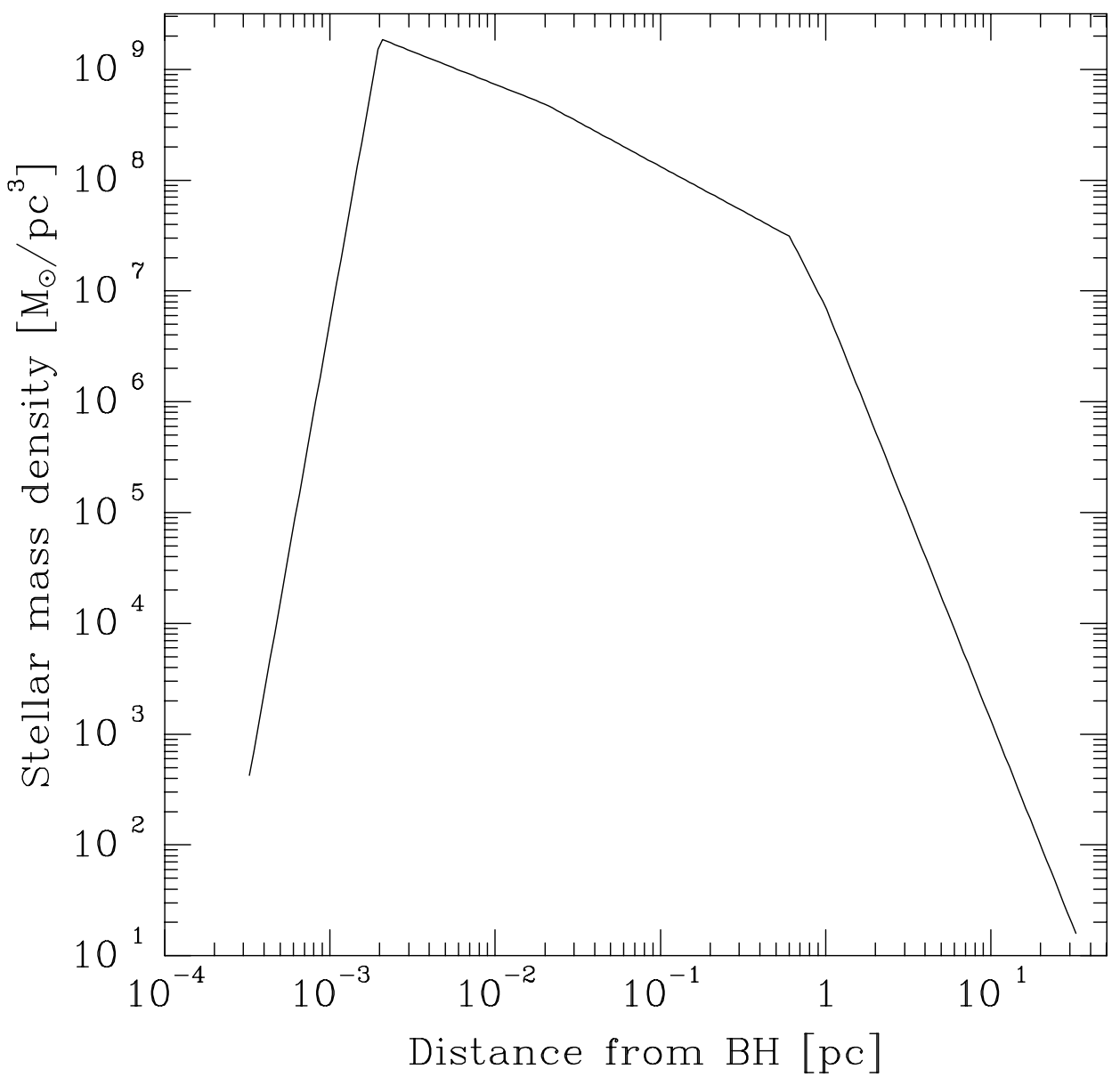

**Figure 1.** The stellar distribution function model.

## 2.2 The bloated star

BSs in the AGN may be very different from solar neighborhood supergiants. Current understanding of stellar winds is based on observations in the solar neighborhood and on theory developed for isolated *spherical* main sequence stars (Mihalas 1978). The wind mechanism of red giants and supergiants is not well understood and furthermore, there are yet no results about the possible effect of the AGN's intense radiation field on the wind structure. A possible mechanism for driving the mass loss, internal radiation pressure on dust formed in the cool regions of the atmosphere (Zuckerman 1980), may be strongly suppressed by the heating of the atmosphere by the external radiation. While the photon flux probably can not increase significantly the mass loss rate (Voit & Shull 1988), a strong enough neutrino flux may transform low mass stars into red giants (MacDonald, Stanev & Biermann 1991) and the existence of a significant neutrino flux in AGN has been suggested (Stecker et al 1991).

In the absence of full reliable models, we adopt the known properties of SGs as the reference point. It is convenient to describe the BS, like a SG, as composed of a star with radius $R_\star$ and mass $M_\star$ which emits from its surface a wind whose denser, line emitting fraction is characterized by a radius $R_{\rm w}$ and mass $M_{\rm w}$.

In the presence of the unidirectional continuum radiation field, the spherical symmetry of an isolated star is replaced by the cylindrical symmetry about the direction of the irradiating flux. For a SG of radius $10^{13}$ cm and a 3000K photosphere, 1/3pc from a $10^{46}$ erg/s continuum source, the external flux at $R_{\rm w} = 10^{14}$cm is an order of magnitude larger than the internal flux. The spherical symmetry must be perturbed and the density and gas velocity of the wind must be strongly affected. In this work we will not attempt to deal with the anisotropy of the wind and shall assume a spherically symmetric wind without specifying the mechanism that is driving the wind. This can be justified if the results of the different wind models show that the line emission is not very sensitive to the details of the wind's internal structure.

The wind's radial profile is obtained from the continuity equation

$$\dot{M}_{\rm w} = 4\pi R^2 v(R) N(R)\ , \qquad (2)$$

where $\dot{M}_{\rm w}$ is the mass loss rate, $R$ is radial distance from the center of the star , $v(R)$ is the wind velocity and $N(R)$ is the gas density. In our models the wind's velocity field is just an alternative way of defining the density structure. Velocity fields which formally describe decelerating flows slower than the escape velocity may be better thought of as describing static envelopes without bulk flow.

Voit & Shull (1988) place lower and upper limits on the SG fraction in the stellar population based on values observed in the solar neighborhood. These limits of $10^{-6}$ and $10^{-3}$ respectively bracket the simple order-of-magnitude estimate of $\sim 10^{-4}$ derived from the mass loss rate of $\sim 10^{-6} M_\odot/$yr and progenitor lifetime of $\sim 3{\times}10^9$ yrs. We shall adopt these solar neighborhood values and assume that $\sim$ 0.01 of the stars in the core are red giants (RG) with $R \sim 10 R_\odot$ and $\dot{M}_{\rm w} \sim 10^{-8} M_\odot/$yr and $\sim 10^{-4}$ of the stars are SGs with $R \sim 100 R_\odot$ and $\dot{M}_{\rm w} \sim 10^{-6} M_\odot/$yr. The rest of the stars are main sequence stars with a typical radius of $1 R_\odot$ and mass-loss rate of $\sim 10^{-14} M_\odot/$yr. If we assume that the gas density at the boundary of the winds is the same for SGs and RGs (see 'Comptonization' below) and that the wind flows at the escape velocity $v(R) = \sqrt{2GM_\star/R}$, then from eq. 2 it follows that the size of the wind scales as $R_{\rm w} \propto \dot{M}_{\rm w}^{\ 2/3}$ and the total geometric cross-section of all SGs is $(10^{-4}/10^{-2}){\times}(10^{-6}/10^{-8})^{4/3} \sim 5$ times larger than that of the RGs. If we take the boundary density as an estimate for the density in the wind, the column density which scales as $R_{\rm w}$ is $\sim$ 20 times larger in a SG. The total mass in the wind scales as $R_w^3$ and therefore the SG winds contain $\sim 100$ times more mass. These ratios justify neglecting the RG winds and main sequence stars and dealing only with the SG winds.

The ionizing radiation may also induce mass loss from the star by radiation pressure at the wind base and by ablation at the star's limbs. Voit & Shull (1988) estimate this contribution to be negligible compared to the normal evolutionary mass loss of supergiants for all stars but those very near to the continuum source. In our model $\dot{M}_{\rm w}$ does not depend on the ionizing flux. We also assume that the BSs fraction in the stellar population is independent of the distance from the black hole.

The ambient HIM pressure may regulate or quench the stellar winds. We assume that the winds expand into vacuum. Further work is needed to clarify this issue, but we note that Voit & Shull (1988) estimate that this assumption is valid in the somewhat different case of X-ray driven winds, 1 pc from the continuum source.

Stellar collisions are important in the dense core and are taken into account in modeling the core evolution (§ 2.1). It is reasonable to expect that the ongoing soft collisions will affect the wind mass and size. We, however, ignore the effects of these repeated collisions on the BS's structure and assume it remains constant throughout its life. This is equivalent to approximating the cumulative effect of the many collisions by a single destructive one.

The radius of the line emitting inner fraction of the wind (henceforth the wind boundary) is an important parameter that determines the covering factor and boundary (i.e. minimal) density of the wind. We consider four physical mechanisms that influence this radius:

(i) Comptonization: Following Kazanas (1989), we assume that all gas with an ionization parameter larger than a certain critical value, corresponding to a minimal boundary density $N_{\rm C}$, goes directly to the hot comptonized phase and its optical depth drops to zero. We define the comptonization radius $R_{\rm C}(r_{\rm bh})$ as the distance where the wind gas density, which decreases with $R$, drops to the comptonization density $N_{\rm C}$ at which point the gas temperature rises to $\sim 5\times10^4$K. It should be emphasized that unlike in the 2-phase model, the cold and hot phases need not be in pressure equilibrium: the wind material is continuously lost to the HIM and replenished by new material from the star's surface. $N_{\rm C}(r_{\rm bh})$, which is a function of continuum flux and depends on the continuum spectrum and chemical composition of the gas, is calculated self-consistently for the model continuum(§ 3.2).

(ii) Finite wind mass: The wind size is limited by the total amount of mass within the wind. We define the average wind mass radius, $R_{\rm mass}$, as the radius within which the average wind mass is contained. The mass loss rate during the super-giant phase of an isolated star is estimated to be (Kudritzki & Reimers 1978)

$$\dot{M}_{\rm w} \sim 5.5\times10^{-13}(L/L_\odot)(R/R_\odot)(M_\odot/M)\ \ M_\odot/yr \ ,(3)$$

with the luminosity estimated at (Paczynski 1970)

$$L = 59250(M_{core}/M_\odot - 0.478)L_\odot \ . \qquad (4)$$

If the SG's mass is small enough it ends its life as a white dwarf with a mass of $\sim 0.6M_\odot$. The mass loss rate is not constant during the super-giant phase but nevertheless we will assume that the line emitting gas can be represented by a wind that contains half of the total ejected mass. The SG mass loss rate of $\dot{M}_{\rm w} \sim 10^{-6}M_\odot/$ yr discussed above is consistent with eqs. 3 and 4 for progenitor stars of $R_\star = 10^{13}$ cm and $1M_\odot$ which have half of the total ejected mass, $0.2M_\odot$, in the wind and $0.8M_\odot$ in the star's core. We use these values for the BSs in all the wind models presented below. A realistic population of SGs includes stars with both less and more than $0.2M_\odot$ in the wind and modeling the emission lines of an average wind mass is not identical to modeling such cases. However, the covering factor is the most important parameter and such a simplification does not alter the overall conclusions.

(iii) Tidal disruption by the black hole: We define the tidal radius as $R_{\rm tidal}(r_{\rm bh}) = X_{\rm tidal}(M_\star/M_{\rm bh})^{1/3}r_{\rm bh}$ where $X_{\rm tidal}$ is a factor of order unity. Since a wind outflowing at velocities above the escape velocity is not gravitationally bound to the star, $R_{\rm tidal}$ should be understood in this context as the radius where the tidal forces disperse the gas effectively enough for immediate comptonization. The results presented below are based on the assumption of $X_{\rm tidal} = 2$.

(iv) Maximal velocity: In some of our models, the wind velocity increases with $R$ indefinitely and an artificial maximal velocity cutoff must be introduced. This maximal velocity depends on the boundary conditions determined by the ambient interstellar gas. A value of 30,000 km/s was chosen on the grounds that this is roughly the largest velocity observed in broadened AGN emission lines. $R_v$ is defined as the radius where the wind velocity reaches the maximum velocity. In practice this constraint is necessary only in extreme cases of highly accelerated wind models.

In all our models, the actual wind boundary radius is taken to be

$$R_{\rm w} = \min(R_{\rm C}(r_{\rm bh}), R_{\rm mass}, R_{\rm tidal}(r_{\rm bh}), R_v) \ . \qquad (5)$$

In certain cases when stars are very near the continuum source the radiation may penetrate the BS's atmosphere. We model the gas density of the star's atmosphere by the simple case of gas in thermal equilibrium in a gravitational potential, giving an exponential density law

$$N_{\rm atm}(R) = N(R_\star)\exp\left(\frac{m_{\rm p}g}{kT}(R_\star - R)\right) \ , \quad R < R_\star \qquad (6)$$

where $m_{\rm p}$ is the proton mass, $T \sim 3000$K and $g = GM_\star/{R_\star}^2$ and can be taken to be constant for $R \sim R_\star$.

Kazanas (1989) suggested that when comptonization determines the wind size, the boundary of the wind is not spherical but has a wide, comet-like tail due to the varying angle of incidence of the irradiating flux relative to the radial pressure of the expanding wind. Comptonization is an atomic process and for a given gas chemical composition and continuum spectrum it is regulated only by the ratio between the photon and electron *densities* which are not vector quantities. For a given wind density structure, the comptonized boundary in the wind hemisphere that faces the continuum is therefore simply a surface of constant density. The boundary surface of the back hemisphere depends on the front-to-back optical depth. In the limbs, where it is negligible, the boundary still follows a constant density surface. At the center, the exact shape of the wind is irrelevant to the line emission as the optical depth is infinite. Between these two limits the back hemisphere's shape may deviate from the surface of constant density. In this work we ignore this possibility and conclude, unlike Kazanas (1989), that a spherically symmetric density structure implies a spherical comptonized boundary (it is possible that a realistic wind density model, such that takes into account the momentum and heat imparted by the continuum, will indeed have an elongated or comet-like comptonized boundary, but this would be a consequence of the non-spherical surfaces of constant density, and not of the flux' angle of incidence).

In the modeling we ignore the possible obscuration of one BS by another. This is consistent with the assumption that the BLR covering is no larger than $\sim 0.1$.

## 3 CALCULATIONS

### 3.1 Photoionization of stellar winds

The photoionization code used in this work is ION (Rees, Netzer & Ferland 1989; Netzer 1993) which implements an iterative algorithm that is based on the approach of treating the atomic physics in detail while approximating the radiative transfer. The treatment of the radiative transfer assumes that the gas has an infinite slab geometry which is homogeneous in density in the transverse plane but may vary in the longitudinal direction. ION receives as input the gas chemical composition, the longitudinal density profile, the external radiation spectral shape and luminosity and produces as output the emission line luminosities. Only "solar composition" models, given by

$$(\mathrm{He}:\mathrm{C}:\mathrm{N}:\mathrm{O}:\mathrm{Ne}:\mathrm{Mg}:\mathrm{Si}:\mathrm{S}:\mathrm{Fe})/\mathrm{H} = 0.0001\times \\ (1000:3.7:1.1:8.0:1.1:0.37:0.35:0.16:0.4) \quad (7)$$

are used in this work (see § 5.1.4).

The application of ION to the stellar winds poses some difficulties. The geometry of the wind is far from slab-like. As will be described in § 3.2, the actual calculations assume a cylindrical shell geometry with a density gradient in both the transverse and longitudinal directions. Another problem particular to stellar winds is the existence of a velocity gradient due to acceleration. For high enough accelerations or lines of relatively small optical depths, the Doppler shift of the line profiles due to the bulk motion of the wind mass may affect resonance scattering and fluorescence. While this can be handled by ION, it was not implemented for the sake of simplicity and because the uncertainties due to the wind structure dominate the results. Various tests that were carried out suggest that two types of lines are most affected. The intensity of the high excitation, temperature sensitive lines, such as O VI $\lambda$1035, decreases with the increase in line width because the reduced optical depth in L$\alpha$ and other major coolants causes a drop in the electron temperature. Intercombination lines, especially C III] $\lambda$1909, can increase their intensity due to the fact that their optical depth is, in some cases, non-negligible. The increased line width reduces the line optical depth thus raising the critical de-excitation density. Both these types of lines are not affected by a large factor.

### 3.2 The numeric procedure

Calculating the line emission from a given core and wind model requires that the line emitting gas be divided into roughly slab-like sections such that can be handled by ION. This proceeds as follows: The stellar distribution is binned into concentric spherical shells. The BSs within a given shell are represented by a single BS whose wind size and density are typical of the shell's mean distance from the black hole. The wind is then divided into concentric annular cylindrical slices whose axis is parallel to the flux of the irradiating continuum. The line emission from each slice is calculated separately by ION given the depth dependent density profile and the results are then integrated back to give the total contribution from the wind of a single BS in the shell. The total line emission from the core is calculated by adding the contributions from each shell, weighted by the number of BSs in it.

The stellar core is cut into a set of radii $\{r_i\}$ between $r_{\rm in}$ and $r_{\rm out}$ such that:

$$\frac{r_{i+1}}{r_i} < r_{\rm ratio} \quad (8)$$

and

$$\frac{\int_{r_i}^{r_{i+1}} r^2 n_\star \, dr}{\int_{r_{\rm in}}^{r_{\rm out}} r^2 n_\star \, dr} < n_{\rm ratio} \quad (9)$$

where $r_{\rm ratio}$ and $n_{\rm ratio}$ are parameters chosen to ensure that the core is finely binned. The values $r_{\rm ratio} = 1.25$ and $n_{\rm ratio} = 1.0$ (i.e. no constraint on the fraction of stars in the shell) were found to give satisfactory results.

Our photoionization calculations show that $r_{\rm in} \sim 10^{16}$ cm is the nearest distance where a sharp transition zone can still be established between the comptonized and cold phases as the high gas density leads to collisional suppression of the coolants. $r_{\rm out}$ is chosen so as to be large enough for the core's integrated line ratios to approach their asymptotic values. The numerical results indicate that $r_{\rm out} = 10^{20}$ cm is a suitable value. The core is typically divided into $\sim 45$ spherical bins.

The wind's density structure is parameterized by the functional form of $v(R)$, which is assumed to be independent of the distance from the black hole, and by the wind size $R_w$. The calculation of $R_w(r)$ from eq. 5 requires the comptonization radius $R_{\rm C}$ and therefore the boundary density $N_{\rm C}$. $N_{\rm C}(r)$ is calculated self-consistently and interpolated from a table prepared beforehand for the specific continuum spectrum used here (§ 5.1.2). The table is prepared for a range of $r_{\rm bh}$ values by applying ION to a gas slab of constant density and by varying this density until the cold phase is established very near the slab's illuminated surface.

The binning of the irradiated cross-section of the wind proceeds by cutting it into a succession of annular slices with radii $\{R_i\}$ from $R = 0$ to $R_w$ such that the slice's area relative to the wind's cross-section satisfies

$$\frac{R_{i+1}^2 - R_i^2}{R_w^2} < A_{\rm ratio} \quad (10)$$

and the maximal wind density along the $i$'th cut, $N_i$, and that of the $i+1$'th cut, $N_{i+1}$ satisfy

$$\frac{N_{i+1}}{N_i} > N_{\rm ratio} \quad (11)$$

where $A_{\rm ratio}$ and $N_{\rm ratio}$ are given parameters. The first constraint is intended to ensure that the binning of the cross-section is fine enough. The second is intended to minimize the transverse density gradient in the gas slices. The values $A_{\rm ratio} = 0.15$ and $n_{\rm ratio} = 0.7$ were adopted by trial-and-error. This choice, which results in very narrow slices (width to depth ratio $< 0.001$ in certain cases) in contradiction with the infinite slab assumption, represents a compromise between conflicting requirements. The number of slices per star depends on the density gradient in the wind and ranges from 7 to $\sim 60$. The luminosity in a given line $\ell$ from a single BS, $L_\ell(r_i)$, is integrated by summing the line flux from each slice weighted by the slice's cross-section. $L_\ell$ is interpolated to any $r_{\rm bh}$ by cubic spline.

**Table 1.** A composite quasar and Seyfert 1 observed line spectrum.

| Line type<br>Line | $F_\ell/F_{\mathrm{L}\alpha}$ range<br>BLR only | $F_\ell/F_{\mathrm{L}\alpha}$ range<br>BLR and NLR |
|---|---|---|
| H and He | | |
| H$\alpha$ $\lambda$6563 | $0.5 \pm 0.3$ | $0.5 \pm 0.3$ |
| H$\beta$ $\lambda$4861 | $0.115 \pm 0.085$ | $0.115 \pm 0.085$ |
| He I $\lambda$5876 | $0.0175 \pm 0.0125$ | $0.0175 \pm 0.0125$ |
| He II $\lambda$1640 | $0.09 \pm 0.06$ | $0.09 \pm 0.06$ |
| Resonance | | |
| C IV $\lambda$1549 | $0.55 \pm 0.25$ | $0.55 \pm 0.25$ |
| N V $\lambda$1240 | $0.225 \pm 0.175$ | $0.225 \pm 0.175$ |
| O VI $\lambda$1035 | $0.375 \pm 0.325$ | $0.375 \pm 0.325$ |
| Very high excitation | | |
| C III $\lambda$977 | $0.05 \pm 0.05$ | $0.05 \pm 0.05$ |
| Intercombination | | |
| C III] $\lambda$1909 | $0.275 \pm 0.125$ | $0.275 \pm 0.125$ |
| Low excitation | | |
| Mg II $\lambda$2798 | $0.275 \pm 0.225$ | $0.275 \pm 0.225$ |
| Forbidden | | |
| [O III] $\lambda$4363 | $0.005 \pm 0.005$ | $0.015 \pm 0.015$ |
| [O III] $\lambda$5007 | $0.0075 \pm 0.0075$ | $0.5 \pm 0.5$ |
| [Ne V] $\lambda$3426 | $0.005 \pm 0.005$ | $0.225 \pm 0.175$ |
| [Fe XI] $\lambda$7892 | $0.005 \pm 0.005$ | $0.01 \pm 0.01$ |

# 4 OBSERVATIONAL AND THEORETICAL CONSTRAINTS

## 4.1 The observed emission line spectrum

We calculate the emission in 15 lines (table 1) as well as the total reprocessed emission in lines and continuum for each of our models. These lines were chosen for the relative reliability in their calculation, the statistical significance of the observed data and their value as probes of the conditions in the gas. The hydrogen and helium lines L$\alpha$, H$\alpha$, H$\beta$, He I $\lambda$5876 and He II $\lambda$1640 are useful for estimating the total absorbed flux and the covering factor. The resonance lines O VI $\lambda$1035, N V $\lambda$1240 and C IV $\lambda$1549 are good indicators of the gas temperature and ionization parameter and the intercombination line C III] $\lambda$1909 is an indicator of intermediate densities of $\sim 10^9\,\mathrm{cm}^{-3}$. The low excitation line Mg II $\lambda$2798 is an indicator of cold, partly ionized gas. The very high excitation line C III $\lambda$977 is an indicator of high density, high temperature gas. The forbidden lines [Ne V] $\lambda$3426, [O III] $\lambda$4363, [O III] $\lambda$5007 and [Fe XI] $\lambda$7892 are indicators of low density gas at various levels of excitation. The O V] $\lambda$1218 line underlying the L$\alpha$ may introduce, in principal, an uncertainty in the observed L$\alpha$ flux. However, since our calculations show that the typical O V] $\lambda$1218/L$\alpha$ ratio in our wind models is less than 1%, we omit this line from the list above.

The calculated line emission from the wind models is expressed in terms of the L$\alpha$ emission and compared to a composite quasar and Seyfert 1 emission line spectrum given in table 1. The range around the mean observed value was chosen so as to contain about 95% of the observed objects. These values are intended for reference only and do not represent a sample mean and standard deviation in the rigorous statistical sense: For some lines the scatter around the mean is actually not symmetric and the 95% range implies in fact a $\pm 2\sigma$ rather than $\pm 1\sigma$ interval. A convenient standard for expressing the discrepancy between a calculated line $\ell$ and the observed range is in terms of the normalized difference, $\hat{L}_\ell = (L_\ell/L_{\mathrm{L}\alpha} - \mu_\ell)/\sigma_\ell$, where $\mu_\ell$ and $\sigma_\ell$ are taken to be the mean and scatter in table 1. In addition to the diagnostics provided by the individual lines, a $\chi^2$ score is calculated with these $\mu_\ell$ and $\sigma_\ell$ values as a measure of the overall goodness-of-fit of all the calculated line ratios. The number of degrees of freedom is $12 = 14 - 2$, the number of independent line ratios minus the number of free parameters in the wind density profile. Since $\hat{L}_\ell$ formally resembles the standard deviation difference, we shall use below terms such as '$3\sigma$ discrepancy', but it should be emphasized that in view of the approximate nature of $\mu_\ell$ and $\sigma_\ell$, the calculated values of $\hat{L}_\ell$ and $\chi^2$ should be interpreted only as rough relative indicators of the models' success.

Reddening is a possible complication in comparing the calculated line ratios to the observed ones. The issue of the intrinsic reddening in AGN is still unclear (MacAlpine 1985). The motivation for assuming some intrinsic reddening is the very large deviation between the observed and calculated hydrogen line ratios, as well as the ratios of weaker lines such as O I and He II (Netzer 1990). In the framework of the screen hypothesis, the application of reddening corrections to these deviations typically gives $E_{\mathrm{B-V}}$ values of 0.05 to 0.25 with a mean of 0.15. Here we use an analytic approximation of the extinction curve based on interstellar observations (Seaton 1979). We correct the results of each wind model, integrated over all the BSs in the core, so as to obtain the mean observed L$\alpha$/H$\beta$ ratio and then correct all the other lines accordingly. In this scheme, similar to that used to deduce the intrinsic reddening in cloud models, $E_{\mathrm{B-V}}$ is not an additional free parameter and may have a different value for each of the wind models. We do not consider the possibility of reddening by internal dust.

## 4.2 The covering factor and fraction of BSs

The BLR covering factor, $C_{\mathrm{F}}$, can be translated into the number of BSs once their line emission is calculated. A possible definition for the covering factor in terms of the AGN's spectral properties is the ratio between the total reprocessed BLR luminosity to the total continuum luminosity, in which case the canonical value is $C_{\mathrm{F}} \sim 0.1$. Alternatively it can be defined in terms of the observed L$\alpha$ equivalent width in which case the canonical value is $EW_{\mathrm{BLR}}(\mathrm{L}\alpha) \sim 100$ Å. We choose the latter since it is more closely related to the quantities directly observed. The required fraction of the of BSs in the stellar population,$f_{\mathrm{BS}}$, is given by

$$f_{\mathrm{BS}} = \frac{100\,\text{Å}}{\int_{r_{\mathrm{in}}}^{r_{\mathrm{out}}} 4\pi r^2 n_\star(r) EW_\star(\mathrm{L}\alpha)(r)\,dr}\,, \qquad (12)$$

where $EW_\star(\mathrm{L}\alpha)$ is the contribution of a single BS to the L$\alpha$ equivalent width and we assume that the fraction of the BSs does not depend on $r_{\rm bh}$. The choice of the integration limits $r_{\rm in}$ and $r_{\rm out}$ will be discussed below in § 4.4. Since $n_\star$ represents only the low mass stars which account for $\sim 1/4$ of the stars in the core (§ 2.1), $f_{\rm BS}$ can take in the extreme case (all the stars are bloated) the maximal value $\sim 4$. If solar neighborhood values apply we expect $f_{\rm BS} \sim 10^{-4}$. Wind models with $f_{\rm BS} \gg 1$ imply inconsistency with the assumed stellar core model.

### 4.3 Mass loss constraints

An important constraint on the number of BSs is mass loss due to stellar evolution in the bloated phase and collisions between BSs. These processes are not part of the MCD calculations which do not assume the existence of BSs. Mass loss from main sequence stars is already included in the $0.3 M_\odot/\,\rm yr$ rate calculated by MCD for our core model (§ 2.1). BS mass loss significantly in excess of this value is inconsistent with the core model and may also imply a large optical depth to electron scattering, $\tau_{\rm es} > 1$, contrary to observations. We will use here a simple model for the mass loss due to collisions between two BSs: The average fractional mass lost at each collision, $\kappa$, is about 0.1 for a collision of two equal mass stars approaching each other with infinite velocity. The fractional mass loss decreases by up to two orders of magnitude for slower collisions and for stars with different masses (Murphy, Cohn & Durisen 1991). We therefore ignore MS–BS collisions and assume $\kappa = 0.1$ for BS–BS and MS–MS collisions. We estimate the velocity dispersion in the core by

$$\sigma_\star(r) = \sqrt{GM(r)/r}\ , \qquad (13)$$

where $M(r)$ is the total mass (black hole and stars) within radius $r$. The collisional mass loss rate per volume due to BS–BS collisions is

$$\dot\rho_{\rm BScoll}(r) = (f_{\rm BS} n_\star(r))^2 \sigma_\star(r) \kappa \pi \times \qquad (14)$$
$$\left[(M_\star - M_{\rm w})R_\star^2 + M_{\rm w}(R_{\rm w}^2(r) - R_\star^2)\right]\ ,$$

where the two different cases of collisions between the BSs themselves and between the winds are considered separately. We also estimate the mass loss due to MS–MS collisions, independently of the MCD result of $0.3 M_\odot/\,\rm yr$, by assuming that the the 1 to 4 overall ratio between $n_\star$ and the total MS stars in the core holds at every $r_{\rm bh}$. Using solar values to represent the MS stars we obtain

$$\dot\rho_{\rm MScoll}(r) = (\max(0, 4 - f_{\rm BS}) n_\star(r))^2 \sigma_\star(r) \kappa \pi M_\odot {R_\odot}^2\ . \qquad (15)$$

The evolutionary mass loss rate per volume from BSs is

$$\dot\rho_{\rm evol} = \dot M_{\rm w} f_{\rm BS} n_\star(r)\ . \qquad (16)$$

These estimates implicitly assume that the ejected mass is spread evenly over the spherical shell containing the star's orbit. The total mass loss rate from these two processes is

$$\dot M_{\rm BS} = \int_{r_{\rm in}}^{r_{\rm out}} 4\pi r^2 (\dot\rho_{\rm BScoll} + \dot\rho_{\rm MScoll} + \dot\rho_{\rm evol})\, dr\ . \qquad (17)$$

On the assumption that this mass is neither accreted nor ejected from the AGN, the upper limit on the buildup rate of the optical depth to electron scattering is

$$\dot\tau_{\rm es} = \sigma_{\rm T} \int_{r_{\rm in}}^{r_{\rm out}} (\dot\rho_{\rm BScoll} + \dot\rho_{\rm MScoll} + \dot\rho_{\rm evol})\, dr\ , \qquad (18)$$

where $\sigma_{\rm T}$ is the Thompson cross-section. Since the BLR size is of the order of at least 0.1 light year, instant removal of this mass is not possible. Even if all this material is eventually accreted, $\dot\tau_{\rm es} \times 1\,\rm yr$ provides an estimate for the mean electron scattering opacity during the phase of the black hole buildup.

### 4.4 BLR size constraints

The integrated luminosities, the number of BSs, the BS fraction in the stellar population, the magnitude of the mass loss effects and the size of the line emitting region all depend on $r_{\rm in}$ and $r_{\rm out}$ and therefore constrain their values. The absence of an intermediate zone between the BLR and NLR may mean that there is some upper cutoff mechanism to the BLR, e.g. dust (Netzer & Laor 1993). While it may be possible that the decreasing line emission efficiency of the BSs will provide a natural explanation for this cutoff, it is also possible that our simplistic assumptions that $\dot M_{\rm w}$ and the fraction of BSs in the population are independent of $r_{\rm bh}$ are wrong and an artificial cutoff must be introduced to correct for it. In this work we calculate the line emission up to a distance of 56 pc (§ 3.2), which is well beyond the estimated BLR size, in order to check this matter explicitly. The exact choice of $r_{\rm in}$ should not affect the the overall results strongly as there are very few stars at the inner radii (although in certain cases their contribution may be significant—see § 5.2.2). In this work we will assume that $r_{\rm in} = 2\times10^{16}$ cm (§ 3.2) and treat $r_{\rm out}$ as yet another free parameter of the model.

As is apparent from eq. 12, $f_{\rm BS}$ is a decreasing function of $r_{\rm out}$. The number of BSs participating in the BLR line emission, $N_{\rm BS}$, is given by

$$N_{\rm BS}(r_{\rm out}) = \frac{100\,\text{Å} \int_{r_{\rm in}}^{r_{\rm out}} r^2 n_\star(r)\, dr}{\int_{r_{\rm in}}^{r_{\rm out}} r^2 n_\star(r) EW_\star({\rm L}\alpha)(r)\, dr} \qquad (19)$$

and is not necessarily a rising function of $r_{\rm out}$.

Observational information on the BLR size obtained from line reverberation campaigns gives a rough estimate of

$$r_{\rm BLR} \sim 0.1 \left(\frac{L}{10^{46}\,\rm erg/s}\right)^{1/2}\ \rm pc\ . \qquad (20)$$

Here we calculate only the flux weighted radius of a line $\ell$

$$r_{\rm fw} = \frac{\int_{r_{\rm in}}^{r_{\rm out}} r^3 n_\star(r) L_\ell(r)\, dr}{\int_{r_{\rm in}}^{r_{\rm out}} r^2 n_\star(r) L_\ell(r)\, dr}\ . \qquad (21)$$

We will not perform a detailed comparison between the flux weighted radius of different lines and limit ourselves to rough consistency checks between the $r_{\rm fw}$ of the prominent BLR lines and $r_{\rm BLR}$ of eq. 20.

## 5 RESULTS

The viability of the bloated stars scenario depends critically on the properties of the single BS as a line emitting object. In presenting the results we aim to separate issues that can be analyzed by studying the properties of the single bloated star from those that depend on the properties of the entire bloated star population in the AGN core. Following that, we try to isolate those features of the bloated stars that do not depend strongly on the choice of wind structure and explore to what extent the structure dependent properties can be varied.

## 5.1 The single bloated star

### *5.1.1 The wind models*

The wind model that was considered in previous works (Kazanas 1989; Begelman & Sikura 1991) was that of a constant velocity flow with $v \sim 10^6$ cm/s. Here we explore a much wider range of wind structures with two forms of velocity profiles

(i) A power-law profile

$$v(R) = v_\star (R/R_\star)^{-\alpha} \tag{22}$$

where $v_\star$ is the velocity at the base of the wind at $R_\star$. This form includes the case of constant velocity, $\alpha = 0$, and of free-fall motion, $\alpha = 0.5$. We explore this form in the range $\alpha = -2.0$ to 0.5 and $v_\star = 5\times10^3$ to $5\times10^6$cm/s. The constant velocity case ($\alpha = 0$) with $v_\star = 10^6$ cm/s will serve below as the reference model.

(ii) An asymptotic profile

$$v(R) = v_\infty(1 - (R/R_\star)^{-\beta}) + v_\star \tag{23}$$

where $v_\infty$ is the velocity at infinity. This analytic approximation reproduces the main features of Voit and Shull's (1988) numeric hydrodynamical work on radiation driven winds. $v_\star = 5\times10^4$ cm/s is the critical velocity at the base of the wind which is assumed to be a constant independent of $v_\infty$ and $\beta$. We explored this form in the range $\beta = 0.5$ to 1.0 and $v_\infty = 5\times10^5$ to $5\times10^6$ cm/s.

Both velocity profiles have two free parameters: $\alpha$ and $v_\star$ for power-law winds and $\beta$ and $v_\infty$ for asymptotic winds.

The asymptotic flow approaches constant velocity at large radii and is therefore expected to produce an emission line spectrum similar to that of a constant velocity wind of the same velocity. Figure 2 presents a comparison of the emission in several representative lines from a constant velocity star of $v = 10^6$ cm/s and and an asymptotic wind flow (eq. 23) with $v_\infty = 10^6$ and $\beta = 0.75$. The boundary density of both wind models, $N_C$, is determined by comptonization and is therefore the same for both at all distances. As expected, at $r_{bh} > 10^{17}$ cm, where $R_w \gg R_\star$, the differences in the line emission relative to L$\alpha$ is no more than 10% and are well within the numeric uncertainty. At smaller distances the constant wind model exhibits emission typical of colder, denser gas relative to the that of the asymptotic wind model: weaker forbidden and semi-forbidden lines, lower C III $\lambda$977and C III] $\lambda$1909emission and higher N V and Mg II emission.

These results can be explained by the difference between the density gradient in the two models: The asymptotic wind gas is denser near $R_\star$ (a consequence of its lower wind base velocity $v_\star = 5\times10^4$ cm/s) and therefore it is comptonized at $R_w > R_\star$ whereas the lower gas density of the constant velocity wind is comptonized at $R_w < R_\star$. The density gradient of the constant velocity model is the steep exponential gradient of the stellar atmosphere which does not allow the existence of an extensive hot, highly ionized zone.

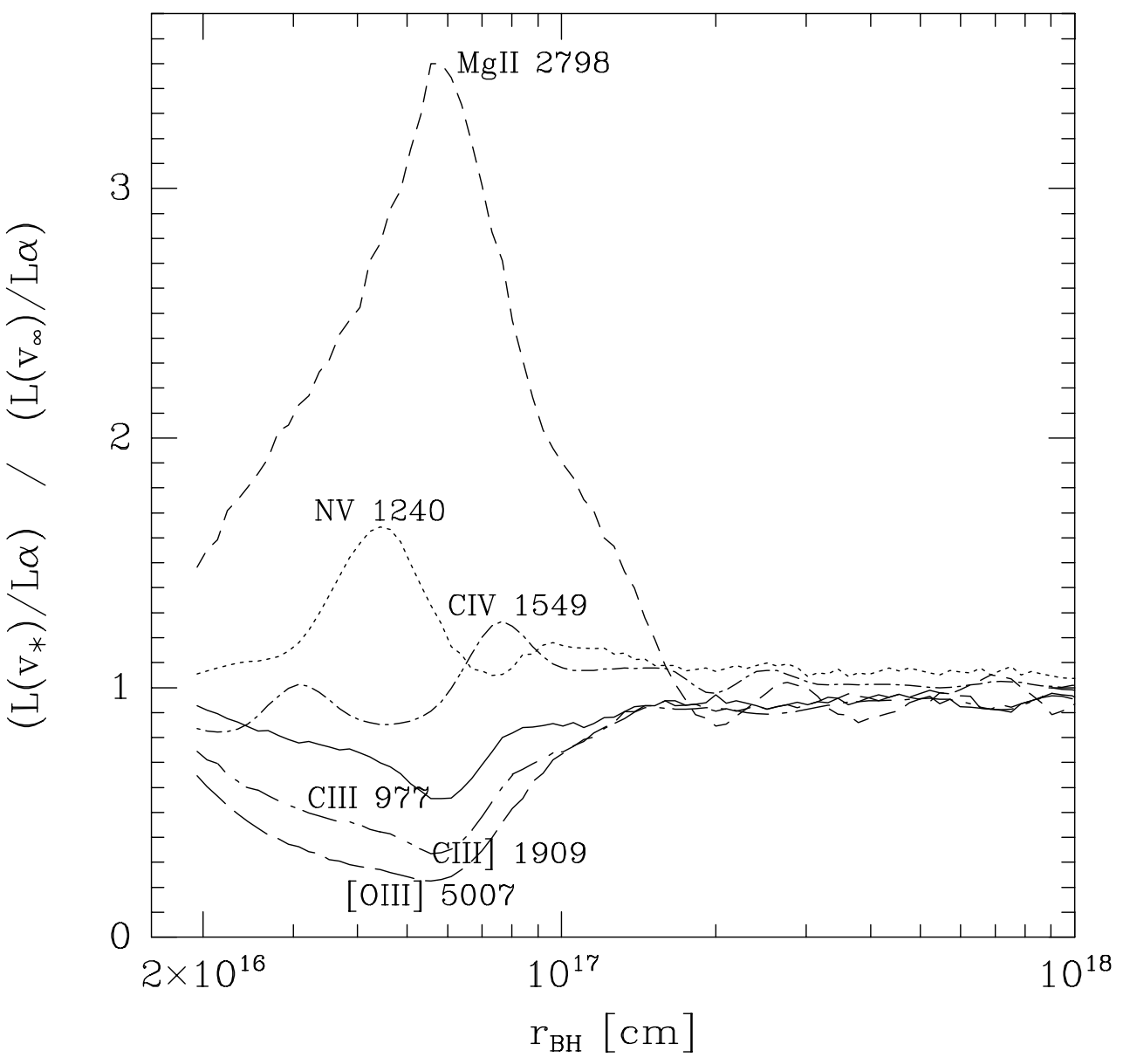

**Figure 2.** Ratios of line emission in two wind models: a constant velocity wind of $v_\star = 10^6$ cm/s and an asymptotic flow wind of $v(R) = 5\times10^4 + 10^6(1-(R/10^{13})^{-0.75})$ cm/s. The ratio plotted is between unreddened luminosities in units of the L$\alpha$ line.

**Table 2.** Boundary density, temperature and ionization parameter as functions of distance from the continuum source.

| $\log(r_{bh}/cm)$ | $\log(N_C/cm^{-3})$ | $T_C$K | $U$ |
|---|---|---|---|
| 16.0 | 12.1 | $7\times10^4$ | 2.8 |
| 16.5 | 11.0 | $7\times10^4$ | 3.5 |
| 17.0 | 10.0 | $7\times10^4$ | 3.5 |
| 17.5 | 9.0 | $7\times10^4$ | 3.5 |
| 18.0 | 8.0 | $7\times10^4$ | 3.5 |
| 18.5 | 7.0 | $8\times10^4$ | 3.5 |
| 19.0 | 6.1 | $9\times10^4$ | 2.8 |
| 19.5 | 5.2 | $9\times10^4$ | 2.2 |
| 20.0 | 4.2 | $5\times10^4$ | 2.2 |
| 20.5 | 3.2 | $5\times10^4$ | 2.2 |
| 21.0 | 2.2 | $5\times10^4$ | 2.2 |
| 21.5 | 1.2 | $5\times10^4$ | 2.2 |

Similar results were also obtained for asymptotic wind models with $\beta = 0.5$ and 1.0 and $v_\infty$ ranging from $5\times10^5$ to $5\times10^6$ cm/s. We conclude that the asymptotic wind models do not differ significantly from constant velocity models in their line emission spectrum and in the following we will concentrate our study on the power-law velocity models.

### *5.1.2 Comptonization and limits on broad forbidden lines*

Under the assumption of comptonization, the outer boundary of the line emitting region of the wind is defined by the onset of instabilities in the cold gas which drive it to the hot phase where it becomes part of the interstellar medium (Kazanas 1989). Unlike previous works, we calculated the instability conditions self-consistently for our assumed continuum. We find that the transition to the unstable branch occurs at a temperature of about 50,000K. The line emission properties of the gas at the comptonization radius are determined by the ionization parameter,

$$U = \frac{\int_{\nu_0}^{\infty} L_\nu / h\nu \, d\nu}{4\pi r^2 c N_{\rm e}} \,. \tag{24}$$

Table 2 displays the dependence of the boundary density, temperature and ionization parameter on the distance from the continuum source. We note that unlike the 2-phase model which predicts that $N_{\rm C} \propto 1/r_{\rm bh}^2$ and that U is constant, our calculations show that $N_{\rm C}$ actually decreases somewhat slower than $1/r_{\rm bh}^2$ and $U$ decreases with the distance. Of greater significance is the fact the values of $U$ at the onset of the stable cold phase are 4–6 times higher than the BLR range of $U = 0.3$ to 1 inferred from photoionization studies and the one implied by the boundary densities assumed by Kazanas (1989).

Gas with ionization parameter values as high as those of our model has the distinct signature of strong very high excitation lines. In particular, we find that some high excitation forbidden lines, with large critical densities (e.g. [Fe XI] $\lambda$7892) can dominate the emission in the outer parts of some winds. Although the wind density is not uniform but increases inwards, the spherical wind is dominated by the thinner limbs. This is a simple consequence of the wind geometry: An infinite slab of thickness $R$ and a thin outer layer of thickness $R - R'$, such as commonly used in numerical modeling of BLR clouds, has only $1 - R'/R$ of its volume occupied by the low density layer, whereas in a sphere of radius $R$ the same layer occupies $1 - (R'/R)^3$ of the volume. As a consequence, the total emission line spectrum of the wind is strongly biased towards the high ionization parameter spectrum (figure 3). Figure 4A shows that the forbidden lines emission from the reference wind model far exceeds the observed BLR values as long as the density at the comptonization radius is lower than the critical density of the forbidden line. In particular, the [Fe XI] $\lambda$7892 emission remains high ($> +3\sigma$) until as close as $\sim 0.2$ pc. Similar results were obtained for all wind models whenever the process that determines the wind boundary is comptonization.

We considered two alternatives for lowering the ionization parameter and thereby for relaxing the constraints on the comptonization mechanism. The first is the assumption of a softer continuum spectrum with mean ionizing photon energy of $\langle h\nu \rangle = 1.27$ Ryd. This spectrum lowers $U$ to 2.4, a value still too high for the BLR, and therefore did not change the forbidden lines luminosities significantly (figure 4B). The second alternative is the assumption of a lower critical temperature of $\sim 3\times10^4$ K. A possible justification may be that the instabilities between the two phases preclude an equilibrium very near the cross-over temperature. This critical temperature lowers $U$ to 0.7–0.9, thereby increasing the collisional suppression distance of the forbidden lines to $\sim 1$ pc, at which point the [Ne V] $\lambda$3426 emission increases above $+3\sigma$ (figure 4C).

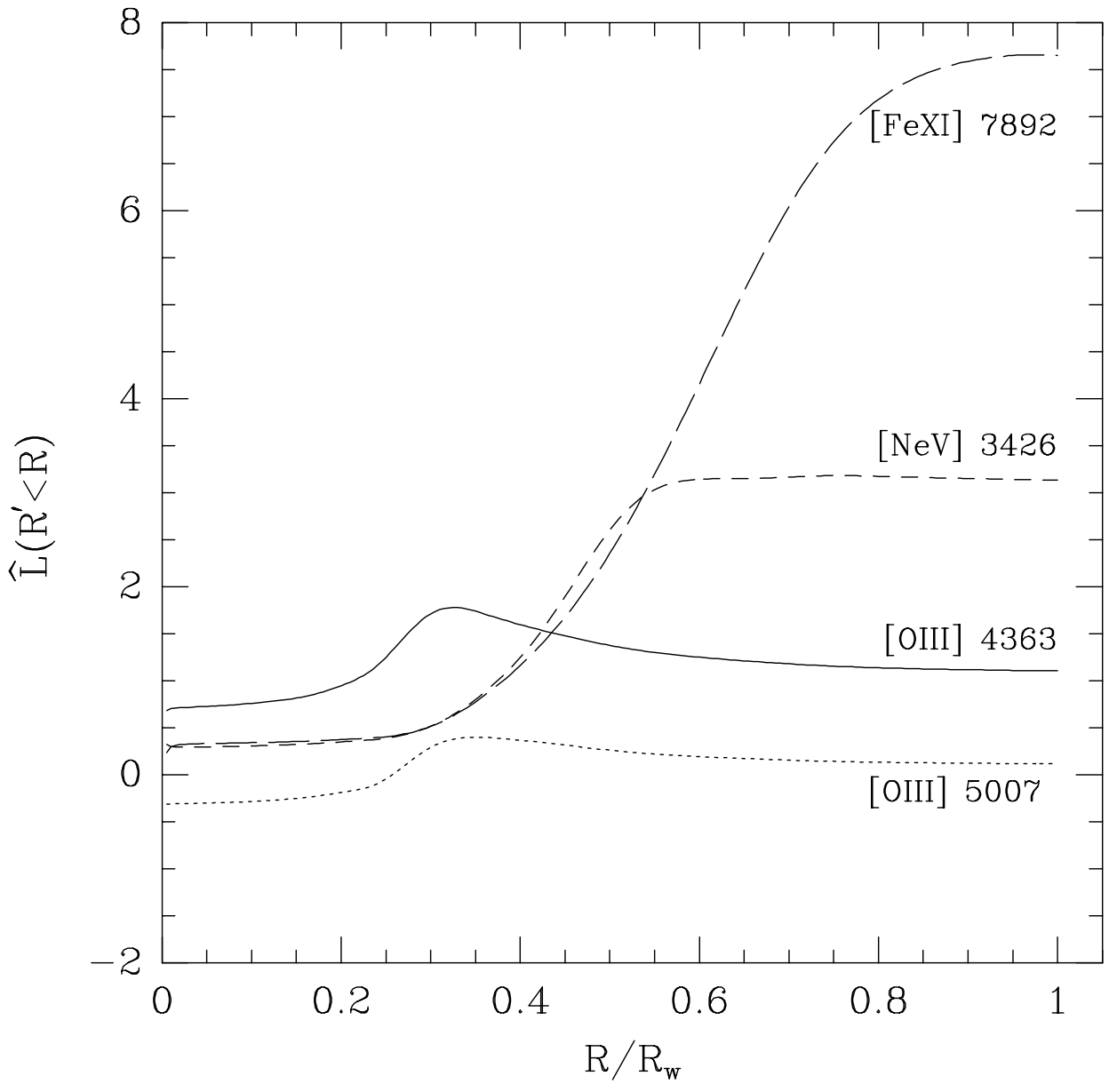

**Figure 3.** Cumulative forbidden line emission from the wind's irradiated cross-section for a star at $r_{\rm bh} = 1$ pc and a constant velocity wind with $v = 10^6$ cm/s. The emission is given in standard deviations from the BLR values of table 1, thus all numbers greater than $\sim 2$ imply excess emission. $R/R_{\rm w} = 0$ is pole-on and 1 is at the outer boundary.

### *5.1.3 Dependence on wind structure*

Table 3 compares the line emission from power-law wind models of different indices and wind base velocities at 0.1 pc and 1 pc. The wind boundary in all models is determined by comptonization and therefore the boundary density, $N_{\rm C}$, is identical in all models at a given distance from the black hole. $N_{\rm C} = 10^9\,{\rm cm}^{-3}$ at 0.1 pc and $N_{\rm C} = 10^7\,{\rm cm}^{-3}$ at 1 pc. The dependence of size of the wind on $\alpha$ and $v_\star$ is obtained from the continuity equation 2:

$$R_{\rm w} = R_\star \left( \frac{\dot M}{4\pi N_{\rm C} v_\star} \right)^{1/(2-\alpha)} = R_\star \left( \frac{N_\star}{N_{\rm C}} \right)^{1/(2-\alpha)} , \tag{25}$$

where $N_\star$ is the gas density at the base of the wind.

The constant velocity models ($\alpha = 0$) have wind base velocities and densities spanning 3 decades, and therefore wind sizes and mean density gradients spanning 1.5 decades. The forbidden lines [O III] $\lambda$4363 and [O III] $\lambda$5007 are the most sensitive to variations of $v_\star$, decreasing by more than a factor of 8 at 0.1 pc and by factors of 5 and 7 respectively at 1 pc as $v_\star$ increases from $5\times10^3$ to $5\times10^6$ cm/s. The semi-forbidden line C III] $\lambda$1909 decreases with velocity by a factor of 5 at 0.1 pc but by less than a factor of 2 at 1 pc. The Mg II $\lambda$2798 line, much lower than its observed value (§ 5.1.4), increases with the velocity by up to a factor of 4 at 0.1 pc but is almost constant at 1 pc. The N V $\lambda$1240 line increases with the velocity by a factor more than 2 at 0.1 pc and by a factor of 3 at 1pc. The C III $\lambda$977 decreases with velocity by more than a factor of 2 at 0.1 and 1 pc. All the other lines change by less than a factor of 2. Some general trends in the results can be readily understood in terms of the changing ratio between the high density and low density gas volumes: The decrease of $R_{\rm w}$ with $v_\star$ lowers the fraction of high density gas in the bloated star with the consequence of decreasing the forbidden and semi-forbidden lines and increasing the Mg II λ2798 line. The greater overall sensitivity of the line spectrum to changes in the velocity at 0.1 pc relative to that at 1 pc has a similar explanation: Although $R_\mathrm{w}$ decreases with the velocity by $10^{1.5}$ at both 0.1 and 1 pc, it is $\sim$ 10 times larger at 1 pc whereas $R_\star$ is constant. The fraction of the high density gas at 1 pc is therefore much smaller and the wind is dominated by the thin outer layers regardless of $v_\star$.

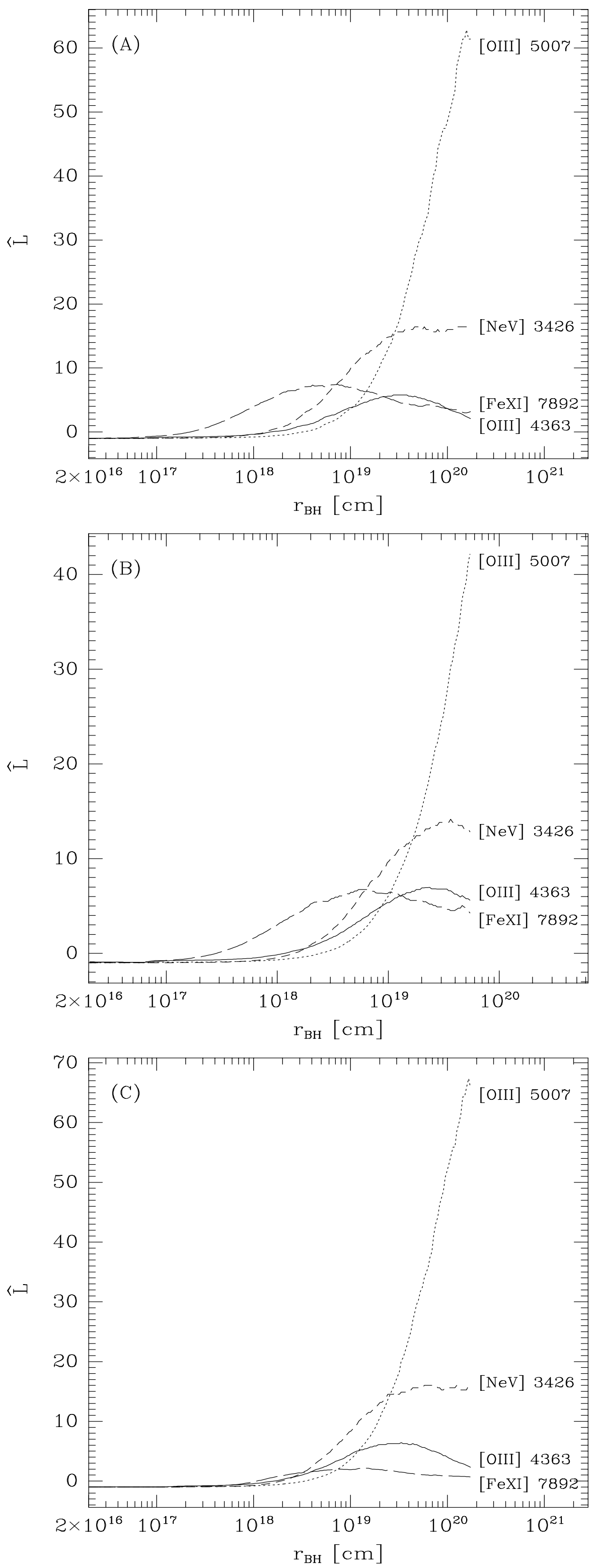

**Figure 4.** Total forbidden line emission from the wind of a single star, $\hat{L}$, as function of distance from the black hole for a constant velocity wind $v = 10^6$ cm/s. The emission is given in standard deviations from the BLR values in table 1. A) Standard ionizing continuum ($\langle h\nu \rangle = 7.74$ Ryd) with a critical boundary temperature of $\sim 6\times10^4$ K. B) Soft continuum of $\langle h\nu \rangle = 1.27$ Ryd with critical boundary temperature of $\sim 6\times10^4$. C) Standard ionizing continuum ($\langle h\nu \rangle = 7.74$ Ryd) with a critical boundary temperature of $3\times10^4$ K.

The accelerated flow models all have the same base velocity of $v_\star = 10^6$ cm/s and base density $N_\star = 3\times10^{10}$ cm$^{-3}$. The dependence of the size of the wind on $N_\mathrm{C}$ (eq. 25) is such that $R_\mathrm{w}$ increases with the wind's power-law index $\alpha$ by a factor of 25 at 1 pc and only by a factor of 4 at 0.1 pc as $\alpha$ increases from $-2.0$ to 0.5. Similarly, $v_\mathrm{C}$, the velocity at the comptonized boundary, decreases with $\alpha$ by a factor of 670 at 1 pc and only by a factor of 16 at 0.1 pc. The mean density gradient in the wind changes by a factor of 7 at 0.1 pc and by a factor of 30 at 1 pc. The variations in the wind structure with $\alpha$ are smaller than that of the constant velocity models, and this is reflected in the behavior of the emission line spectrum. Although the line spectrum changes with decreasing wind radius similarly to that of the constant velocity models, the magnitude of the changes is smaller and is more pronounced at 1 pc rather than at 0.1 pc due to the much greater variation range. The forbidden lines [O III] λ4363 and [O III] λ5007 increase with $\alpha$ by a factor of 2.5 at 0.1 pc and by factors of 4 and 6 respectively at 1 pc. The forbidden [Ne V] λ3426 line increases with $\alpha$ by more than a factor of 2 at 0.1 pc and by a factor of 4 at 1 pc. The forbidden [Fe XI] λ7892 which changes by less than a factor of 2 at 0.1 pc decreases with $\alpha$ by a factor of 4 at 1 pc. The Mg II λ2798 line increases with $\alpha$ by a factor of more than 2 at 0.1 pc and 3 at 1 pc. The O VI λ1035 line is almost independent of $\alpha$ at 0.1 pc but increases with $\alpha$ by a factor of 3 at 1 pc. All other lines change by less than a factor of 2.

### *5.1.4 The* Mg II λ2798/Lα *and* N V λ1240/Lα *deficiency*

The two line ratios Mg II λ2798/Lα and N V λ1240/Lα are too low compared to the observed BLR values regardless of the wind model parameters or the distance from the black hole. The Mg II λ2798/Lα deficiency is the most conspicuous discrepancy. This discrepancy is exhibited by all the power-law wind models at distances less than $\sim$ 1 pc from the black hole and up to distances of $\sim$ 20 pc by all winds but those of dense and slow decelerating flows. Reddening can improve the ratio only by a factor of $\sim$ 2 even for the extreme case of $E_{\mathrm{B-V}} = 0.2$. Figure 5 shows the the range of the Mg II/Lα ratios emitted by a single bloated star from power-law wind models covering the full parameter range as function of the distance from the black hole. The Mg II/Lα ratio is less than 0.03 ($\hat{L} = -1.1$) for all wind models up to $\sim$ 1 pc and less than 0.07 ($\hat{L} = -0.9$) up to 56 pc for all but the densest and slowest wind models.

The Mg II is an indicator of cold, partly neutral gas. The Mg II emission pattern reflects an interplay between the wind temperature and column density: At small distances from the black hole, a significant fraction of the wind cross-section is occupied by the bloated star itself which can provide the

**Table 3.** The dependence of the line emission spectrum of a single BS on the power-law wind parameters.

| | $\alpha = 0$ | | $r_{\rm bh} = 0.1$ pc | |
|---|---|---|---|---|
| $v_\star$ cm/s | 5(3) | 5(4) | 5(5) | 5(6) |
| $R_{\rm w}$ cm | 7.6(14) | 2.4(14) | 7.6(13) | 2.4(13) |
| $N_\star$ cm$^{-3}$ | 6(12) | 6(11) | 6(10) | 6(9) |
| H$\alpha$ $\lambda$6563 | 1.6(−1) | 9.4(−2) | 8.7(−2) | 8.9(−2) |
| H$\beta$ $\lambda$4861 | 5.7(−2) | 3.7(−2) | 3.3(−2) | 3.1(−2) |
| He I $\lambda$5876 | 4.4(−3) | 4.2(−3) | 3.9(−3) | 5.7(−3) |
| He II $\lambda$1640 | 4.2(−2) | 4.5(−2) | 5.7(−2) | 7.2(−2) |
| He II $\lambda$1640 | 2.8(−1) | 2.9(−1) | 3.3(−1) | 3.5(−1) |
| N V $\lambda$1240 | 2.5(−2) | 2.6(−2) | 3.5(−2) | 5.5(−2) |
| O VI $\lambda$1035 | 4.1(−1) | 4.2(−1) | 5.0(−1) | 6.4(−1) |
| C III $\lambda$977 | 7.8(−2) | 6.8(−2) | 5.6(−2) | 3.2(−2) |
| C III] $\lambda$1909 | 3.9(−1) | 3.1(−1) | 2.1(−1) | 8.2(−2) |
| Mg II $\lambda$2798 | 2.2(−3) | 2.3(−3) | 2.8(−3) | 8.8(−3) |
| [O III] $\lambda$4363 | 4.9(−3) | 3.0(−3) | 1.6(−3) | 5.6(−4) |
| [O III] $\lambda$5007 | 2.3(−3) | 1.4(−3) | 7.7(−4) | 2.8(−4) |
| [Ne V] $\lambda$3426 | 6.2(−4) | 5.6(−4) | 5.3(−4) | 4.5(−4) |
| [Fe XI] $\lambda$7892 | 6.5(−3) | 6.5(−3) | 7.3(−3) | 8.7(−3) |

| | $\alpha = 0$ | | $r_{\rm bh} = 1$ pc | |
|---|---|---|---|---|
| $v_\star$ cm/s | 5(3) | 5(4) | 5(5) | 5(6) |
| $R_{\rm w}$ cm | 7.1(15) | 2.2(15) | 7.1(14) | 2.2(14) |
| $N_\star$ cm$^{-3}$ | 6(12) | 6(11) | 6(10) | 6(9) |
| H$\alpha$ $\lambda$6563 | 1.4(−1) | 1.2(−1) | 1.1(−1) | 1.0(−1) |
| H$\beta$ $\lambda$4861 | 1.8(−2) | 1.9(−2) | 1.9(−2) | 2.0(−2) |
| He I $\lambda$5876 | 3.3(−3) | 3.1(−3) | 3.1(−3) | 3.2(−3) |
| He II $\lambda$1640 | 3.9(−2) | 4.9(−2) | 6.0(−2) | 7.1(−2) |
| C IV $\lambda$1549 | 3.3(−1) | 3.6(−1) | 4.1(−1) | 4.8(−1) |
| N V $\lambda$1240 | 2.3(−2) | 3.3(−2) | 4.9(−2) | 6.7(−2) |
| O VI $\lambda$1035 | 3.2(−1) | 4.1(−1) | 5.2(−1) | 5.7(−1) |
| C III $\lambda$977 | 3.0(−2) | 2.5(−2) | 1.9(−2) | 1.6(−2) |
| C III] $\lambda$1909 | 2.1(−1) | 1.8(−1) | 1.4(−1) | 1.1(−1) |
| Mg II $\lambda$2798 | 9.3(−3) | 9.0(−3) | 1.0(−2) | 1.1(−2) |
| [O III] $\lambda$4363 | 2.8(−2) | 1.8(−2) | 1.0(−2) | 6.0(−3) |
| [O III] $\lambda$5007 | 2.8(−2) | 1.6(−2) | 7.6(−3) | 4.0(−3) |
| [Ne V] $\lambda$3426 | 1.9(−2) | 1.8(−2) | 1.5(−2) | 1.4(−2) |
| [Fe XI] $\lambda$7892 | 2.0(−2) | 3.0(−2) | 3.8(−2) | 3.8(−2) |

| | $v_\star = 10^6$ cm/s | | $r_{\rm bh} = 0.1$ pc | |
|---|---|---|---|---|
| $\alpha$ | 0.5 | 0.0 | −1.0 | −2.0 |
| $R_{\rm w}$ cm | 9.5(13) | 5.4(13) | 3.1(13) | 2.3(13) |
| $v_{\rm C}$ cm/s | 3.2(5) | 1.0(6) | 3.1(6) | 5.3(6) |
| H$\alpha$ $\lambda$6563 | 8.3(−2) | 8.4(−2) | 8.7(−2) | 8.6(−2) |
| H$\beta$ $\lambda$4861 | 3.2(−2) | 3.1(−2) | 3.1(−2) | 3.0(−2) |
| He I $\lambda$5876 | 3.8(−3) | 3.9(−3) | 4.5(−3) | 5.0(−3) |
| He II $\lambda$1640 | 6.0(−2) | 6.3(−2) | 5.8(−2) | 5.5(−2) |
| C IV $\lambda$1549 | 3.4(−1) | 3.4(−1) | 3.4(−1) | 3.4(−1) |
| N V $\lambda$1240 | 3.5(−2) | 4.0(−2) | 4.5(−2) | 4.6(−2) |
| O VI $\lambda$1035 | 5.2(−1) | 5.3(−1) | 4.9(−1) | 4.6(−1) |
| C III $\lambda$977 | 5.3(−2) | 5.3(−2) | 5.4(−2) | 5.5(−2) |
| C III] $\lambda$1909 | 2.2(−1) | 1.9(−1) | 1.8(−1) | 1.6(−1) |
| Mg II $\lambda$2798 | 2.4(−3) | 3.0(−3) | 4.8(−3) | 6.7(−3) |
| [O III] $\lambda$4363 | 1.9(−3) | 1.3(−3) | 9.7(−4) | 7.8(−4) |
| [O III] $\lambda$5007 | 8.9(−4) | 6.3(−4) | 4.7(−4) | 3.8(−4) |
| [Ne V] $\lambda$3426 | 6.5(−4) | 5.2(−4) | 3.6(−4) | 2.9(−4) |
| [Fe XI] $\lambda$7892 | 7.9(−3) | 7.6(−3) | 6.0(−3) | 4.8(−3) |

| | $v_\star = 10^6$ cm/s | | $r_{\rm bh} = 1$ pc | |
|---|---|---|---|---|
| $\alpha$ | 0.5 | 0.0 | −1.0 | −2.0 |
| $R_{\rm w}$ cm | 1.8(15) | 5.0(14) | 1.4(14) | 7.1(13) |
| $v_{\rm C}$ cm/s | 7.5(4) | 1.0(6) | 1.4(7) | 5.0(7) |
| H$\alpha$ $\lambda$6563 | 1.1(−1) | 1.1(−1) | 1.2(−1) | 1.2(−1) |
| H$\beta$ $\lambda$4861 | 1.9(−2) | 1.9(−2) | 1.8(−2) | 1.8(−2) |
| He I $\lambda$5876 | 2.9(−3) | 3.1(−3) | 3.7(−3) | 4.1(−3) |
| He II $\lambda$1640 | 6.2(−2) | 6.2(−2) | 4.9(−2) | 4.1(−2) |
| C IV $\lambda$1549 | 3.7(−1) | 4.2(−1) | 5.2(−1) | 5.3(−1) |
| N V $\lambda$1240 | 4.2(−2) | 5.3(−2) | 5.1(−2) | 4.2(−2) |
| O VI $\lambda$1035 | 5.3(−1) | 5.2(−1) | 3.1(−1) | 2.0(−1) |
| C III $\lambda$977 | 2.0(−2) | 1.9(−2) | 2.0(−2) | 2.1(−2) |
| C III] $\lambda$1909 | 1.3(−1) | 1.3(−1) | 1.5(−1) | 1.6(−1) |
| Mg II $\lambda$2798 | 7.2(−3) | 1.0(−2) | 1.7(−2) | 2.3(−2) |
| [O III] $\lambda$4363 | 1.5(−2) | 9.1(−3) | 5.5(−3) | 3.9(−3) |
| [O III] $\lambda$5007 | 1.3(−2) | 6.5(−3) | 3.3(−3) | 2.2(−3) |
| [Ne V] $\lambda$3426 | 2.2(−2) | 1.5(−2) | 8.3(−3) | 5.1(−3) |
| [Fe XI] $\lambda$7892 | 4.0(−2) | 3.8(−2) | 1.7(−2) | 8.1(−3) |

Top left panel: constant velocity winds with different wind base velocities at 0.1 pc. Top right panel: as in top left panel, at 1 pc. Bottom left panel: accelerated flow winds with the same wind base velocity at 0.1 pc. Bottom right panel: as in bottom left panel, at 1 pc. The line luminosities are given relative to L$\alpha$, unreddened.

high column densities required for cold, partly ionized gas. However, the low ionization is attained at high column densities where the optical depth of Mg II is very large. At larger distances, where the wind is dominated by the low density outflowing gas and $\tau$ is low, the relatively high ionization parameter maintains the gas at a state where the magnesium is ionized several times. The combination of low ionization *and* low $\tau$ is found only in the very slow and dense wind models, whose size, and therefore boundary density, are determined by the mass constraint $R_{\rm mass}$ (Fig. 6) rather than by comptonization. This results in a much higher boundary density and consequently a much lower ionization parameter and low gas temperature.

We investigated whether it is possible to increase the Mg II emission in the comptonized reference wind model by using the softer continuum or by lowering the boundary temperature as in § 5.1.2. this resulted in an increase of less than a factor of 2 in the former and less than a factor of 3 in the latter and therefore does not change the general conclusion that the comptonized wind models show a Mg II deficiency.

The deficiency in the N V $\lambda$1240/L$\alpha$ ratio is another persistent feature of the results. Again, this deficiency cannot be explained by reddening. Adopting a lower comptonization temperature increased the ratio only at $r_{\rm bh} < 0.1$ pc by less than 20%. The softer continuum luminosity increased this ratio by only $\sim 20\%$ at all radii which is still very low compared to the observed values.

The N V $\lambda$1240/L$\alpha$ problem is related to the general composition problem discussed recently by Hamann & Ferland (1993). These authors suggest that the abundance of nitrogen and other elements in high luminosity, high redshift quasars is much larger than their solar abundances, causing a significant increase in the above ratio. If this is indeed the case, our solar composition models should not be compared to the average N V $\lambda$1240/L$\alpha$ ratio, which is affected by the inclusion of such quasars in the sample. We did not investigate high metalicity models and leave this possible discrepancy for future work.

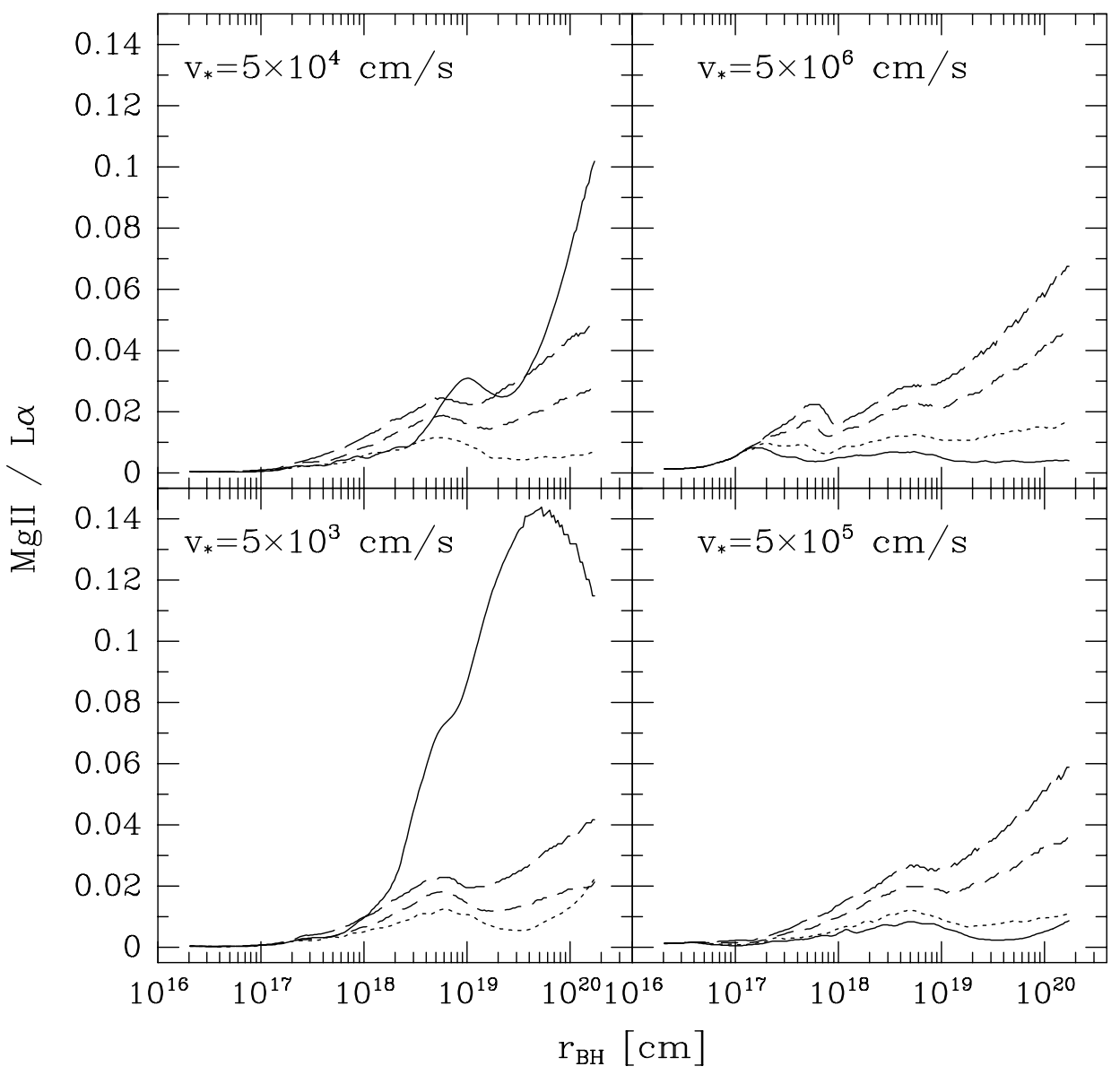

**Figure 5.** The Mg II$(r_{\rm bh})$ / L$\alpha(r_{\rm bh})$ luminosity ratio from power-law wind models. Solid line: $\alpha = 0.5$, dotted line: $\alpha = 0.0$, short dashed line: $\alpha = -1.0$, long dashed line: $\alpha = -2.0$.

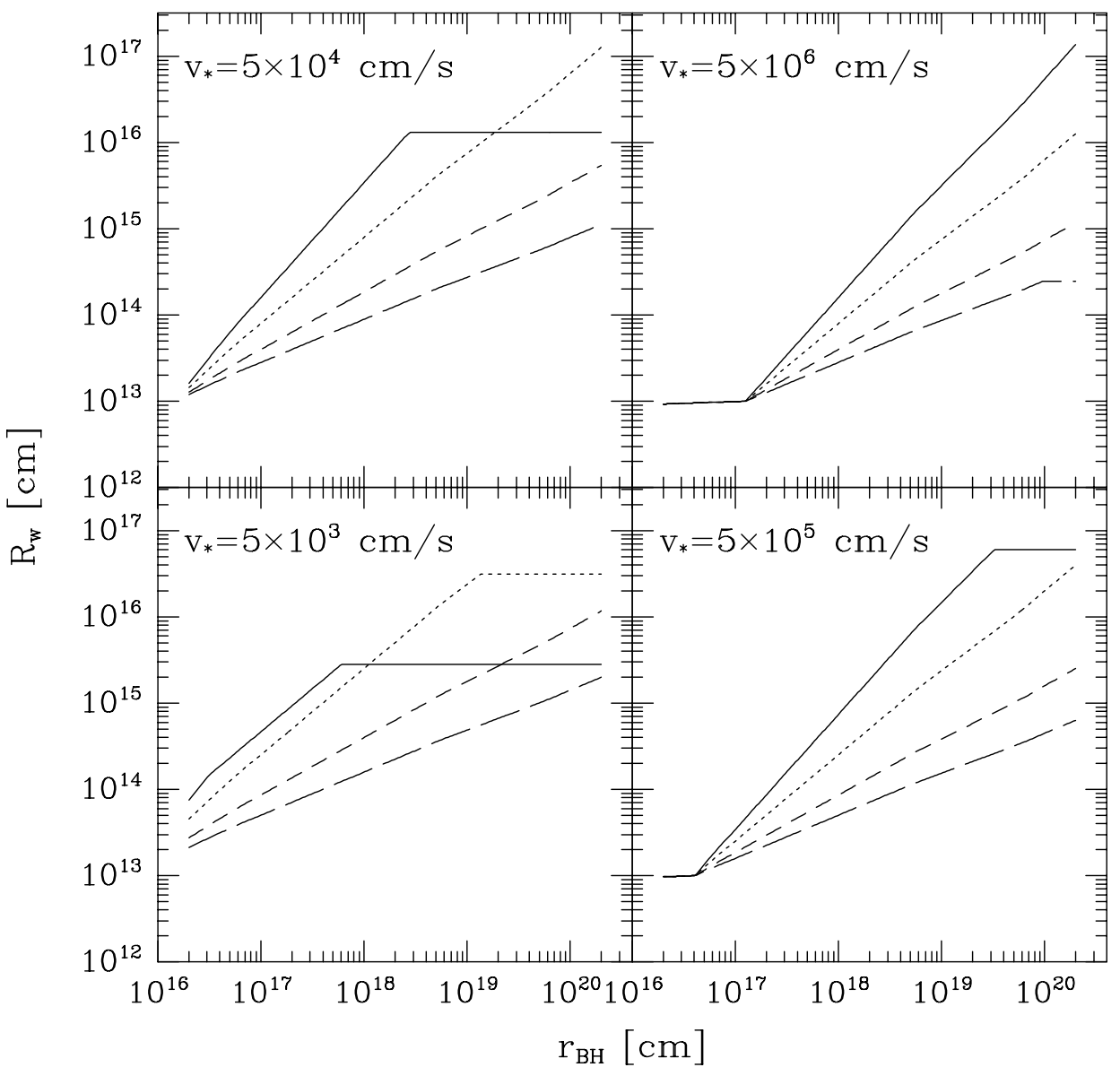

**Figure 6.** Wind sizes for power-law wind models. The wind boundary is determined by comptonization subject to the constraints of $X_{\rm tidal} = 2$, $M_{\rm w} \leq 0.2 M_\odot$ and $v_{\rm max} = 0.1\,$c. Solid line: $\alpha = 0.5$, dotted line: $\alpha = 0.0$, short dashed line: $\alpha = -1.0$, long dashed line: $\alpha = -2.0$.

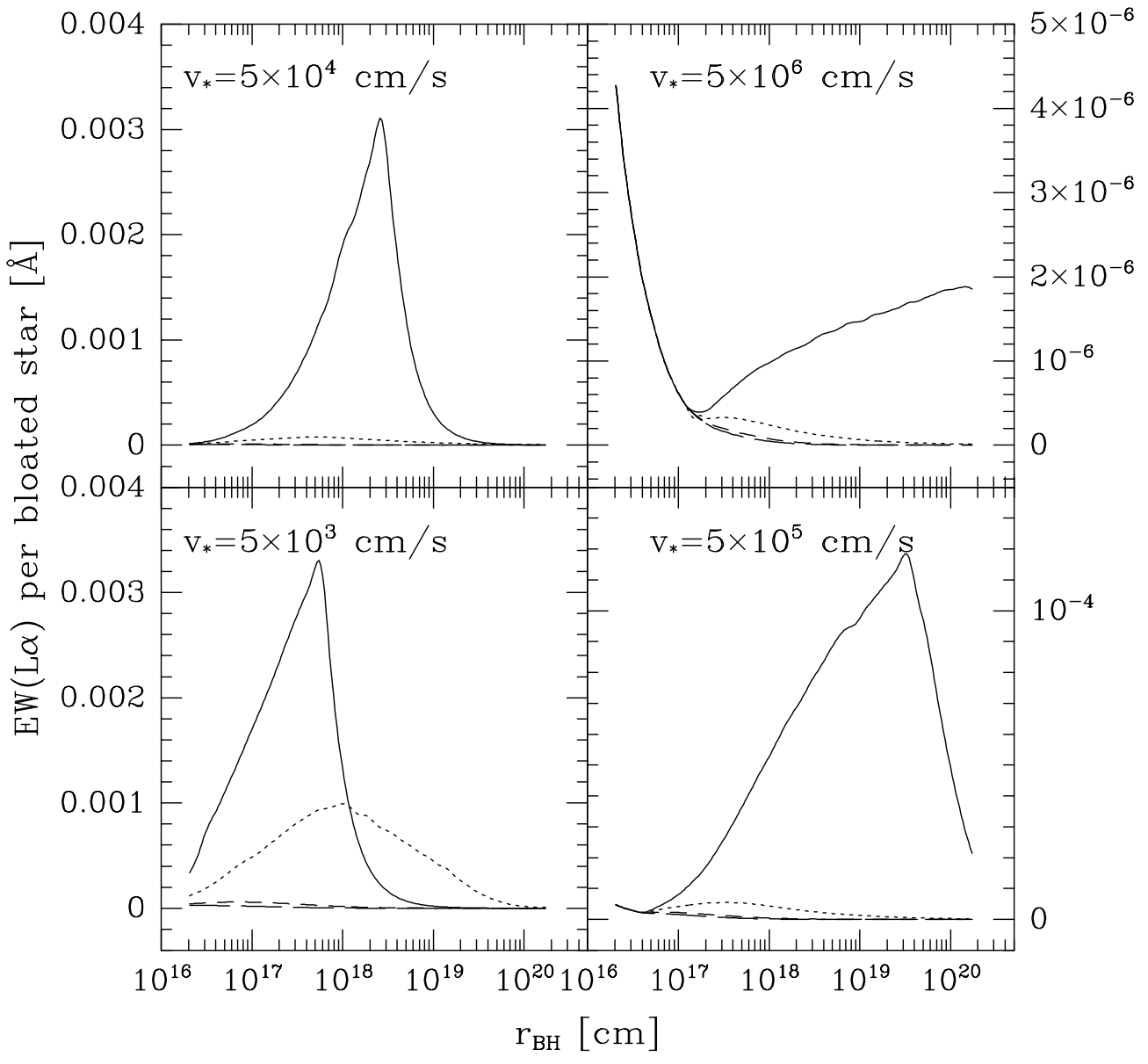

**Figure 7.** $EW$(L$\alpha$) per bloated star as function of the distance from the black hole. Solid line: $\alpha = 0.5$, dotted line: $\alpha = 0.0$, short dashed line: $\alpha = -1.0$, long dashed line: $\alpha = -2.0$.

### *5.1.5 Line emission efficiency of bloated stars*

The line emission efficiency of the bloated star depends on both its size and density structure. Figure 6 shows the wind sizes of the power-law models as function of the distance from the black hole. The wind size is calculated from eq. 5. The tidal disruption constraint affects only the slowest and densest wind model of $\alpha = 0.5$, $v_\star = 5{\times}10^3$ cm/s and the maximal velocity constraint only the fastest and most accelerated wind model of $\alpha = -2.0$ and $v_\star = 5{\times}10^6$. The size of the slow wind models is determined at the larger distances by the constraint on the mass in the wind and the denser the wind the earlier the onset of the wind mass constraint. As $r_{\rm bh}$ decreases, the sizes of the winds decrease until they reach $R_\star = 10^{13}$ cm, the size of the bloated star itself. From that point on the wind size remains nearly constant due to the very steep exponential density profile of the stellar atmosphere. The distance at which $R_{\rm w} = R_\star$ decreases as the wind density increases.

Figure 7 shows the contribution of a single star to the L$\alpha$ equivalent width as function of the distance from the black hole. This contribution is less than $10^{-4}$ Å at all distances for all but the slow and dense models. The peaks of the contribution coincide with the distance where the wind size stops increasing with $R_{\rm C}$ and is constrained by $R_{\rm mass}$. From that point on the irradiating flux decreases as $1/r_{\rm bh}^2$ and the contribution drops steeply.

Previous estimates of the wind's actual line emission efficiency were based on the assumption that the wind's geometric cross-section, or a constant fraction thereof, is optically thick. Figure 8A shows the ratio between the total reprocessed luminosity from the wind to the total ionizing luminosity that is irradiating the wind's geometrical cross-section $\pi R_w^2$. At small distances the ratio exceeds 1 due to the large column densities of the bloated star itself which also absorb photons redwards of the Lyman limit. As $r_{\rm bh}$ increases a larger fraction of the outer part of the wind becomes optically thin. The ratio does not descend smoothly with $r_{\rm bh}$ but displays various bumps and local maxima, especially at small $r_{\rm bh}$. This reflects the existence of ionization from excited states which can be significant at small distances where the gas is highly excited. We note that all but the dense and slow wind models exhibit a similar efficiency which decreases from $\sim 1$ at $10^{17}$ cm to $\sim 0.1$ at $10^{20}$ cm. The slow and dense wind models have a consistently higher actual to geometric covering factor with some showing a sharp rise at large distances. This merely expresses the fact that the mass cutoff on the wind size excludes the thinner gas that does not contribute significantly to the optical depth of the wind.

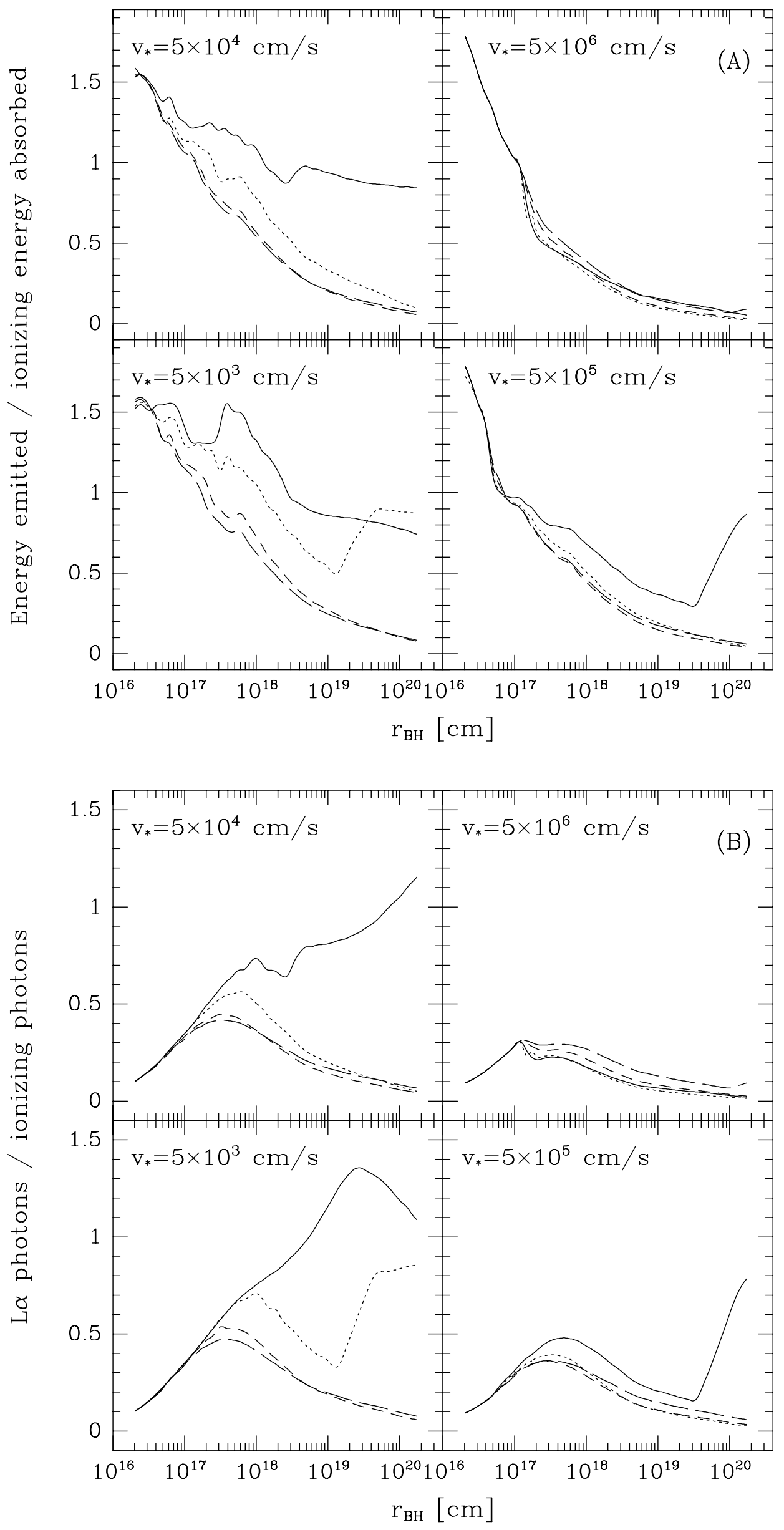

**Figure 8.** Effective covering factor relative to the geometrical cross-section of the power-law wind models as function of the distance from the black hole. A) Expressed in the ratio of reprocessed to irradiating luminosity. B) Expressed in the ratio of L$\alpha$ photons to infalling irradiating ionizing photons. Solid line: $\alpha = 0.5$, dotted line: $\alpha = 0.0$, short dashed line: $\alpha = -1.0$, long dashed line: $\alpha = -2.0$.

Another way of expressing the relation between the geometrical cross-section and the efficiency of the winds is by the ratio of the emitted L$\alpha$ photons to the the number of irradiating ionizing photons. This is displayed in figure 8B which shows that at small distances, where the winds are very dense, the high densities and large optical depths suppress the L$\alpha$. Again, all the models but those of slow and dense winds display a similar behavior and the slow and dense models are more efficient in converting ionizing photons into L$\alpha$ and exhibit a rise in this efficiency at large distances.

## 5.2 The bloated star population

### *5.2.1 The integrated line emission*

The line emission was integrated with the cutoff radii 0.1 pc, 1/3 pc, 1 pc and 56 pc. Figure 9 shows the overall reduced $\chi^2$ goodness-of-fit score for the integrated line emission with an extinction coefficient $E_{\rm B-V}$ that is required for correcting the L$\alpha$/H$\beta$ ratio for $r_{\rm out} = 1/3$ pc and 56 pc. The results for the 0.1pc and 1pc cutoffs display the same trends as those of the 1/3pc with the fit becoming worse as $r_{\rm out}$ increases due mainly to the over-production of the forbidden lines. The extinction coefficient is in the range $E_{\rm B-V} \sim 0.20$ to 0.25 for all the models but the $\alpha = 0.5$, $v_\star = 5\times10^3$ cm/s model which has the low $E_{\rm B-V} = 0.03$. In almost all the wind models the reddening correction worsens the $\chi^2$ score, an effect due mostly to the increase in the reddened forbidden lines.

Figure 10A compares the integrated line emission from the stellar core to the observed BLR values for the reference wind model as function of the cutoff radius. This wind model illustrates trends which are common to all the wind models: The low H$\alpha$, H$\beta$ and He I (before reddening correction), the decrease of the high density and temperature C III $\lambda$977 line with increasing $r_{\rm out}$ and the increase in the forbidden line emission with increasing $r_{\rm out}$ (The forbidden line emission at high cutoffs is outside the graph's scale and is comparable to NLR values). The forbidden lines aside, the variation of the emission line spectrum with $r_{\rm out}$ is not very large. Figure 10B shows the integrated line ratios for the slow and dense wind model $\alpha = 0.5$, $v_\star = 5\times10^3$ cm/s. The line ratios, apart of the persistent deficiency of Mg II $\lambda$2798 and N V $\lambda$1240, fit the observed ones quite well.

### *5.2.2 Line emission from the innermost BSs*

Figure 11 shows the differential line emission of several lines and of the total reprocessed luminosity emitted from a stellar core containing BSs with the reference model's wind structure. The emission exhibits a decrease at small radii up to $\sim 6\times10^{16}$ cm, then an increase up to $2\times10^{18}$ cm followed by a sharp decline at larger radii. The relatively large contribution of the very few innermost BSs can be explained by an interplay between the two density regimes of the BS—the wind and the star's atmosphere—and the core's stellar density distribution function. For optically thick winds the total heat reprocessed by a spherical shell of BSs is given by the product of the irradiating flux, the wind's cross-section and the number of BSs in the shell

$$dH(r_{\rm bh}) \propto \frac{L}{4\pi r_{\rm bh}^2} \times \pi R_w^2 \times 4\pi r_{\rm bh}^2 n_\star(r_{\rm bh}) dr \,. \qquad (26)$$

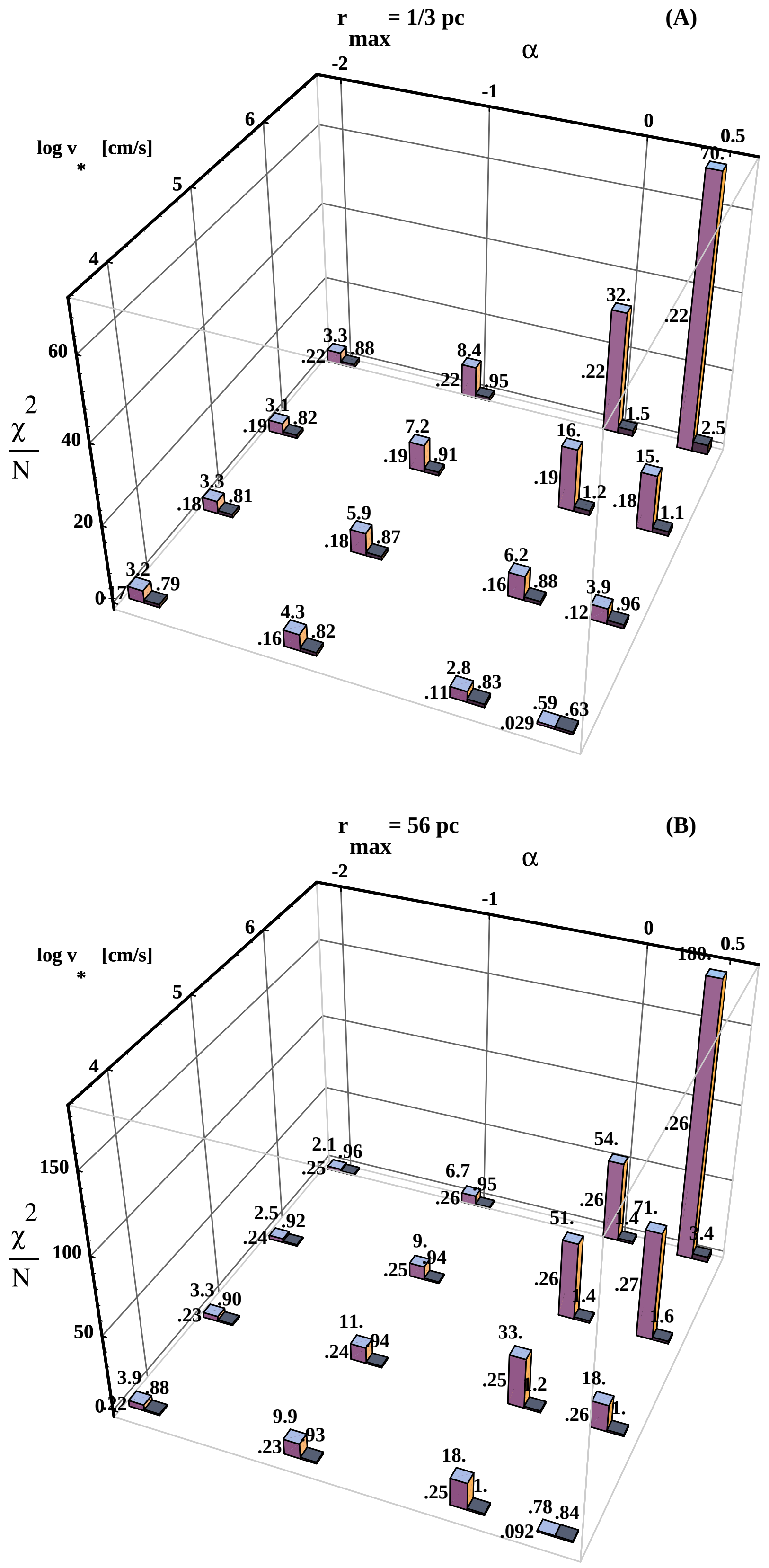

**Figure 9.** The reduced $\chi^2$ goodness-of-fit score and $E_{\mathrm{B-V}}$. The dark gray column on the right represents the unreddened value, the light gray column on the left, the reddened value. The numeric values of the $\chi^2$ are given above the columns. $E_{\mathrm{B-V}}$ is given on the left of each pair of columns. A) $r_{\mathrm{out}} = 1/3\,\mathrm{pc}$. $\chi^2$ based on the BLR values. B) $r_{\mathrm{out}} = 56\,\mathrm{pc}$. $\chi^2$ based on the BLR and NLR values.

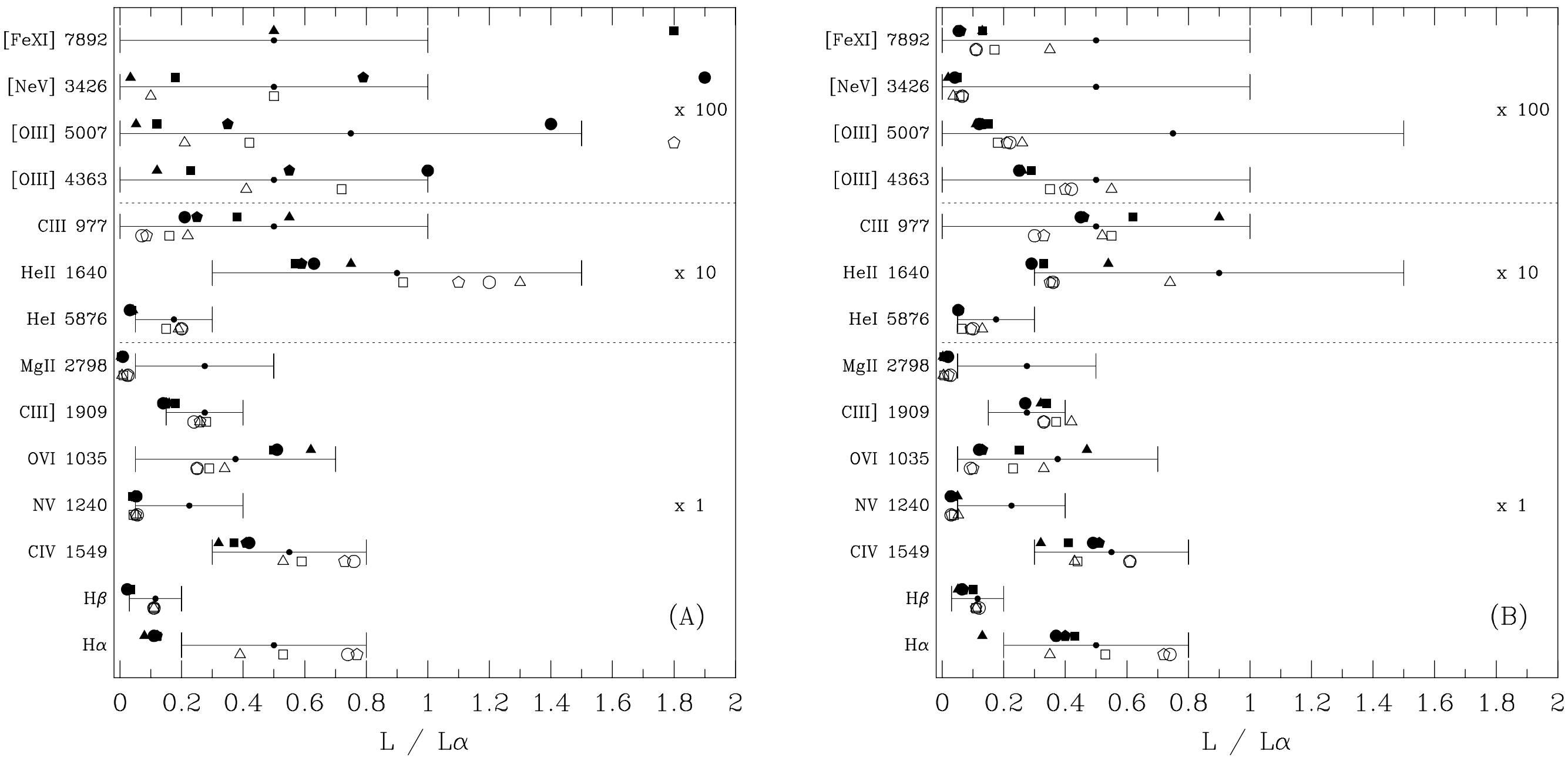

**Figure 10.** The integrated line emission (relative to Lα) from the stellar core compared to the observed BLR values. The mean observed value is given by the small dot and the range by the error bars. The black triangle, square, pentagon and circle above the error bars give the unreddened ratio for cutoffs at 0.1 pc, 1/3 pc, 1 pc and 56 pc respectively. The white shapes beneath the error bar give the ratios after being corrected for reddening. A) The reference wind model $\alpha = 0.0$, $v_\star = 10^6$ cm/s. $E_{\rm B-V}(0.1\,{\rm pc}) = 0.22$, $E_{\rm B-V}(1/3\,{\rm pc}) = 0.20$, $E_{\rm B-V}(1\,{\rm pc}) = 0.25$, $E_{\rm B-V}(56\,{\rm pc}) = 0.26$. B) The slow and dense wind model $\alpha = 0.5$, $v_\star = 5{\times}10^3$ cm/s. $E_{\rm B-V}(0.1\,{\rm pc}) = 0.13$, $E_{\rm B-V}(1/3\,{\rm pc}) = 0.03$, $E_{\rm B-V}(1\,{\rm pc}) = 0.08$, $E_{\rm B-V}(56\,{\rm pc}) = 0.09$.

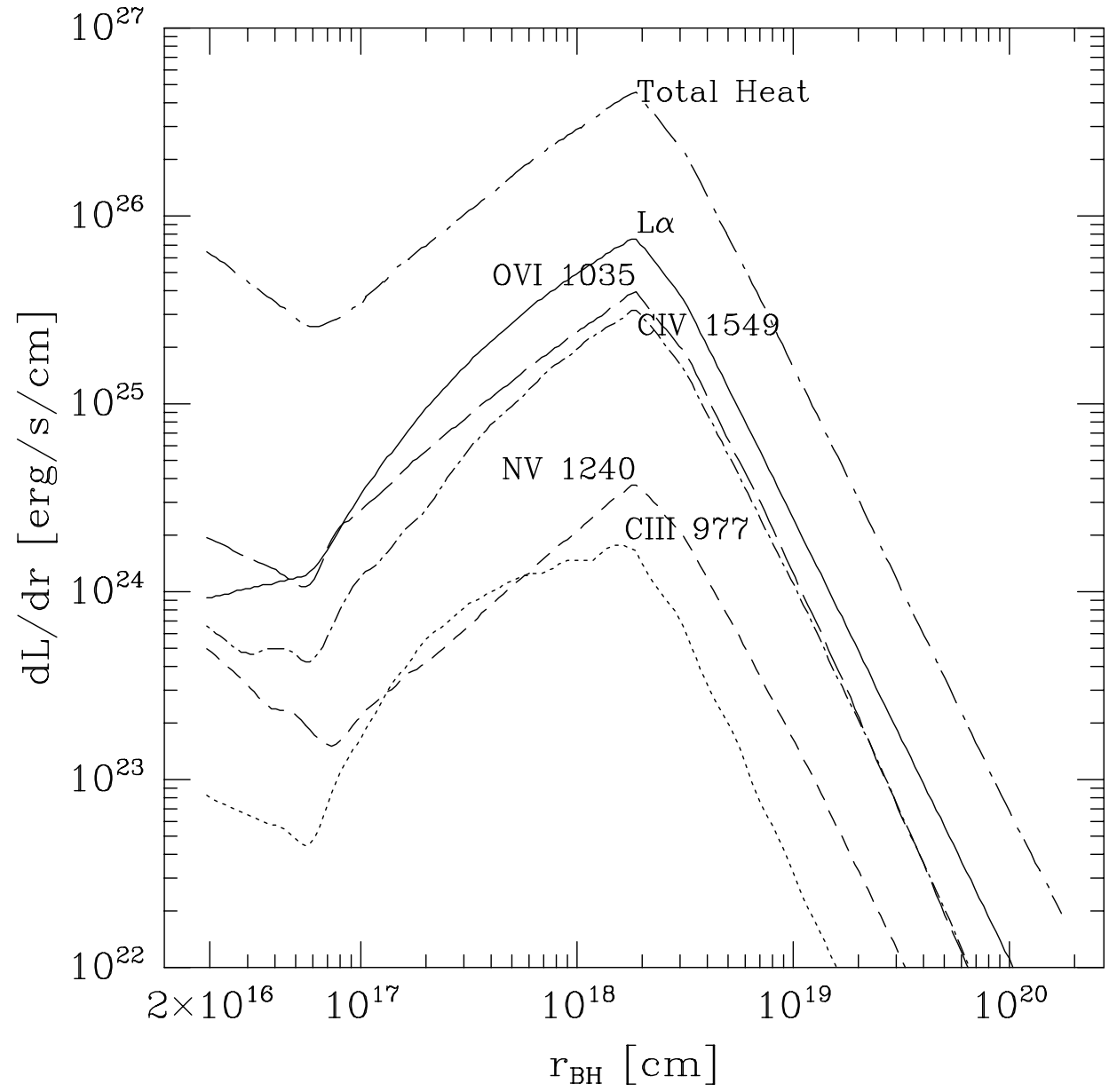

**Figure 11.** Differential line emission as function of the distance from the black hole for BSs with constant velocity wind of $v = 10^6$ cm/s.

At small radii where $R_{\rm w} = R_{\rm C} < R_\star$, $R_{\rm w}$ is almost constant (see fig. 6) due to the steep exponential density gradient of the star's atmosphere and therefore $dH/dr \propto n_\star(r_{\rm bh})$. For the reference wind model this occurs at $r_{\rm bh} < 6{\times}10^{16}$ cm. The stellar distribution $n_\star$ increases with decreasing radius until it reaches a peak at $6{\times}10^{15}$ cm $\sim 2.5$ light days. Although our calculations extend only to as close as $2{\times}10^{16}$ cm from the black hole, we expect that the differential total heat emission, following $n_\star$, will peak at 2.5 light days and then decline sharply. If we assume that the line luminosity approximately follows that of the total heat and extrapolate linearly to the peak we find that the differential luminosity in the lines C III λ977, C IV λ1549 and O VI λ1035 at 2.5 light days is greater than that of all radii up to $\sim 10^{17}$ cm, and for N V λ1240, even up to $\sim 10^{18}$ cm.

The distance where $R_{\rm w} = R_{\rm C}$ depends on the wind density through the parameter $\dot{M}_{\rm w}/v_\star$: The denser the wind, the smaller this distance is. At larger radii, where $R_{\rm w} > R_\star$, the size of a comptonized constant velocity wind such as the reference model, is proportional to $N_{\rm C}^{1/2} \propto r_{\rm bh}$ and therefore $dH/dr \propto 4\pi r_{\rm bh}^2 n_\star$. The sharp turn-over at $2{\times}10^{18}$ cm reflects the point where $n_\star$ begins to fall off faster than $r_{\rm bh}^2$. The decreasing optical thickness of the wind also contributes to the decline in $dH/dr$. While realistically not all the wind's cross-section is optically thick and $R_{\rm w}$ is not strictly constant at small radii, this qualitative explanation does describe the overall behavior of the detailed photoionization results.

We note that unlike the resonance lines and the high excitation C III λ977 line, the Lα emission does not increase at smaller radii. This is a consequence of the increasing op-tical depth and the line thermalization. It results in large line to L$\alpha$ ratios in the innermost BSs.

#### *5.2.3 The number and mass loss of BSs*

Figure 12 shows the fraction of BSs in the core's stellar population and their number for the various models and the 1/3pc and 56pc cutoffs. The results reflect the line emission efficiency of the individual wind structures. The highest $f_{\rm BS}$ values are needed for the fast, highly accelerating wind models and the lowest for the slow, decelerating models. These trends are also exhibited by the 0.1pc and 1pc cutoffs. We note that the $\alpha = 0.0$, $v_\star = 5\times10^6$ cm/s model, very similar to the reference model, requires a high ($> 4$) fraction of BSs at all cutoff radii.

Figure 13 shows the total mass loss rate and the buildup rate of the electron scattering optical depth. The fast and accelerated flow models with $\alpha \leq -1$ or $v_\star > 5\times10^5$ cm/s all show very high ($> 10 M_\odot/$ yr) mass loss rates at all cutoffs. At the smallest cutoff, 0.1 pc, even the densest wind model is associated with rather high mass loss and optical depth buildup rates of $\dot{M} = 6.5 M_\odot/$ yr and $\dot{\tau} = 0.013\,{\rm yr}^{-1}$.

The striking feature in these results is the large difference between the properties of the different wind models: At a given cutoff radius, $f_{\rm BS}$ extends over a range of 4 orders of magnitude, $\dot{M}$ over a range of 6 orders of magnitude and $\dot{\tau}_{\rm es}$ over a range of 5 orders of magnitude.

#### *5.2.4 The BLR size*

Figure 14 shows the flux weighted radius, $r_{\rm fw}({\rm H}\beta)$, for the various models with a cutoff of $r_{\rm out} = 56$ pc. The flux weighted radii are much smaller than 56 pc since the stellar density $n_\star$ falls rapidly beyond 1 pc. The general trends in the behavior of $r_{\rm fw}({\rm H}\beta)$ as a function of the wind model parameters $\alpha$ and $v_\star$ are also exhibited by all the other lines, at all cutoffs. With the exception of the two slow decelerating models $\alpha = 0.5$, $v_\star = 5\times10^3$ and $5\times10^4$ cm/s, there is a slow increase in $r_{\rm fw}$ with $v_\star$ and a somewhat steeper rise with $\alpha$. This reflects the $\alpha$ and $v_\star$ dependence of the wind size $R_{\rm w}(r_{\rm bh})$ and the line emission efficiency (figures 6 and 7). The small flux weighted radius of the two slow wind models is a consequence of their sharply peaked line emission efficiency which reaches its maximum at small radii due to the $R_{\rm mass}$ cutoff on the wind size. $r_{\rm fw}({\rm H}\beta)$ is, up to a factor of 2, also the flux weighted radius of the other lines with the exception of the forbidden lines (especially the [O III] $\lambda$5007 line) which are emitted at much larger distances. We note that the $\alpha = 0.5$, $v_\star = 5\times10^3$ cm/s model does not require any cutoff to be consistent with $r_{\rm BLR}$ of eq. 20.

At cutoffs $\leq 1$ pc, $r_{\rm fw}$ of the lines is roughly proportional to the cutoff radius itself. At these small cutoffs the difference in $r_{\rm fw}$ of the various lines is small. Depending on the wind model and the specific line, the flux weighted radii range from a 1/3 to 2/3 of $r_{\rm out}$.

## 6 DISCUSSION

### 6.1 The scope of this work

Previous investigations of the bloated stars scenario did not follow the path of detailed line emission calculations, but rather relied on various simplifying assumptions to predict the emission line spectrum (Penston 1988; Kazanas 1989; Scoville & Norman 1988). By assuming normal stellar evolution and the properties of supergiants, or by estimating the effects of the irradiating flux on the star, they proceeded to check different aspects of the model: The effects of the radiation on the star, its size and mass loss rate (Mathews 1983; Scoville & Norman 1988; Voit & Shull 1988; Kazanas 1989), the role of the stars in feeding the black hole and the luminosity evolution (Norman & Scoville 1988), the correlation between the luminosity and covering factor (Kazanas 1989), the line profiles (Scoville & Norman 1988; Kazanas 1989) and the implication of the collisional rate (Penston 1988; Scoville & Norman 1988; Begelman & Sikura 1991). These works did not treat the atomic physics and line emission processes in detail and considered only constant velocity winds, with properties similar to those of normal supergiants. In particular they did not take account of the varying line emitting efficiency of the winds and did not employ a detailed, self-consistent model of the stellar core.

Here we adopt a different approach. The AGN environment is drastically different from normal stellar environments and its effects on stars and the course of stellar evolution are currently unknown. This is to be contrasted with the atomic physics involved in the line emission which, although computationally complex, are well understood theoretically and well tested against observations in various astrophysical domains. We therefore use the line emission as the basis of our investigation. The use of a photoionization code allows us to predict correctly the effects of the ionizing continuum on the wind: The comptonization of the outer layers, the line emitting efficiency and the emitted line spectrum. The known properties of normal supergiants are used only as a reference point for a wide range of wind density structures. These wind models are then evaluated according to their success in reproducing the observed line properties, with no a-priori prejudice as to the 'reasonable' wind model. We use a stellar core model based on detailed numeric calculations (Murphy, Cohn & Durisen 1991). This allows us to assign the proper weight to the line emission from different parts of the core, to check for consistency with the total number of stars in the core and to estimate the magnitude of the collisional processes once we obtain the number of the bloated stars from the photoionization calculations.

The various components and assumptions of the stellar core / bloated star model were discussed in detail in § 2. The main input parameters of our models are the choice of the stellar core's mass and evolutionary stage, the continuum luminosity spectrum and the BS's evolutionary mass loss rate, wind mass and wind density structure. All other physical quantities are derived from these parameters. We studied the line emission properties of the single BS as a function of the wind structure and distance from the black hole with particular emphasis comptonization and line emission efficiency. The integrated line emission from all the BSs in the core was compared to the observed spectrum. We in-vestigated the constraints placed on the various wind models by the number of BSs and their collisional mass loss rates. In addition we studied the possibility of a significant contribution from the relatively few BS very near the black hole and compared the BLR size, as measured by the flux weighted radii of the lines, to that derived from line reverberation studies.

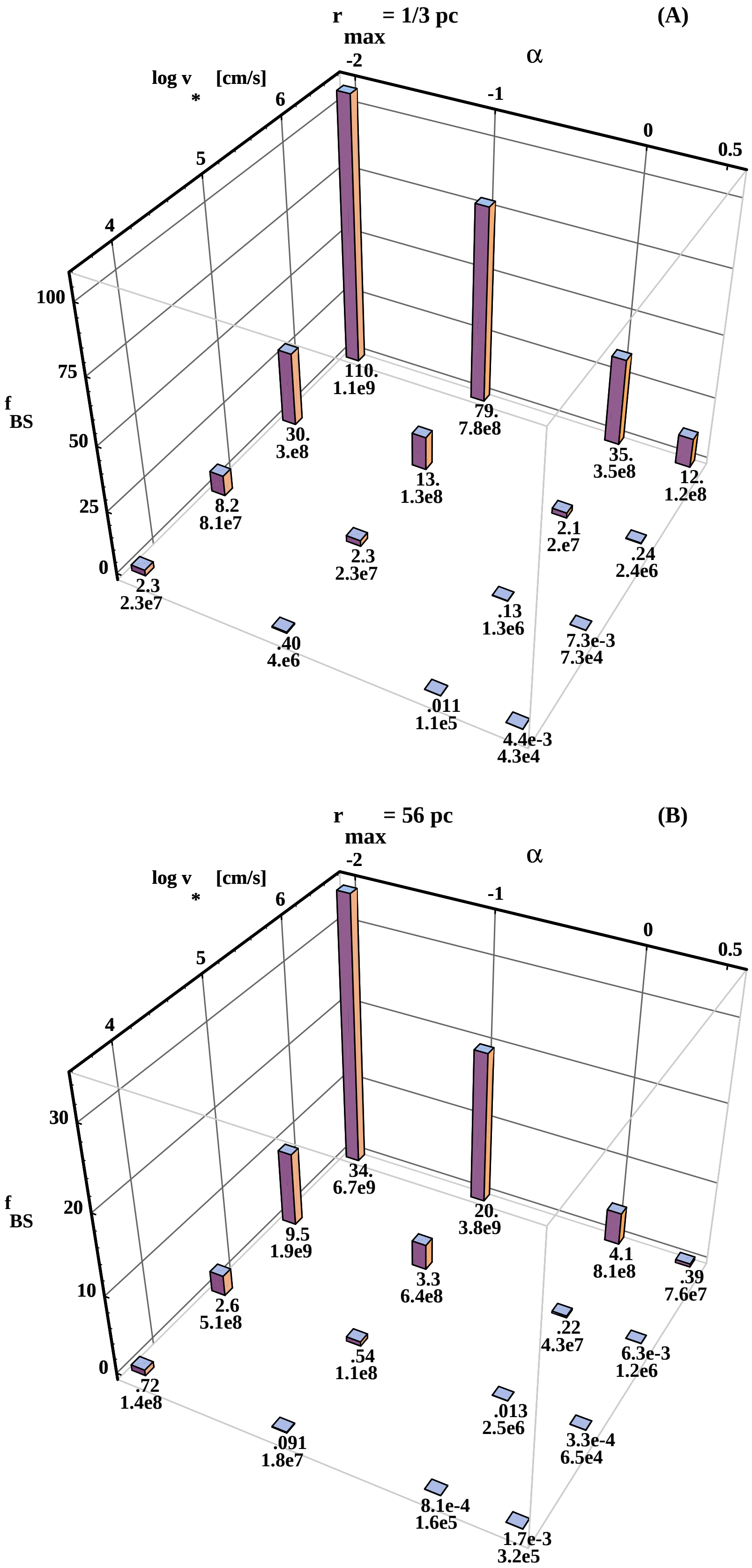

**Figure 12.** The fraction, $f_{\rm BS}$, and number, $N_{\rm BS}$, of the BSs in the stellar population. The top number at the foot of each column is $f_{\rm BS}$ and the bottom one is $N_{\rm BS}$. A) $r_{\rm out} = 1/3\,{\rm pc}$. B) $r_{\rm out} = 56\,{\rm pc}$.

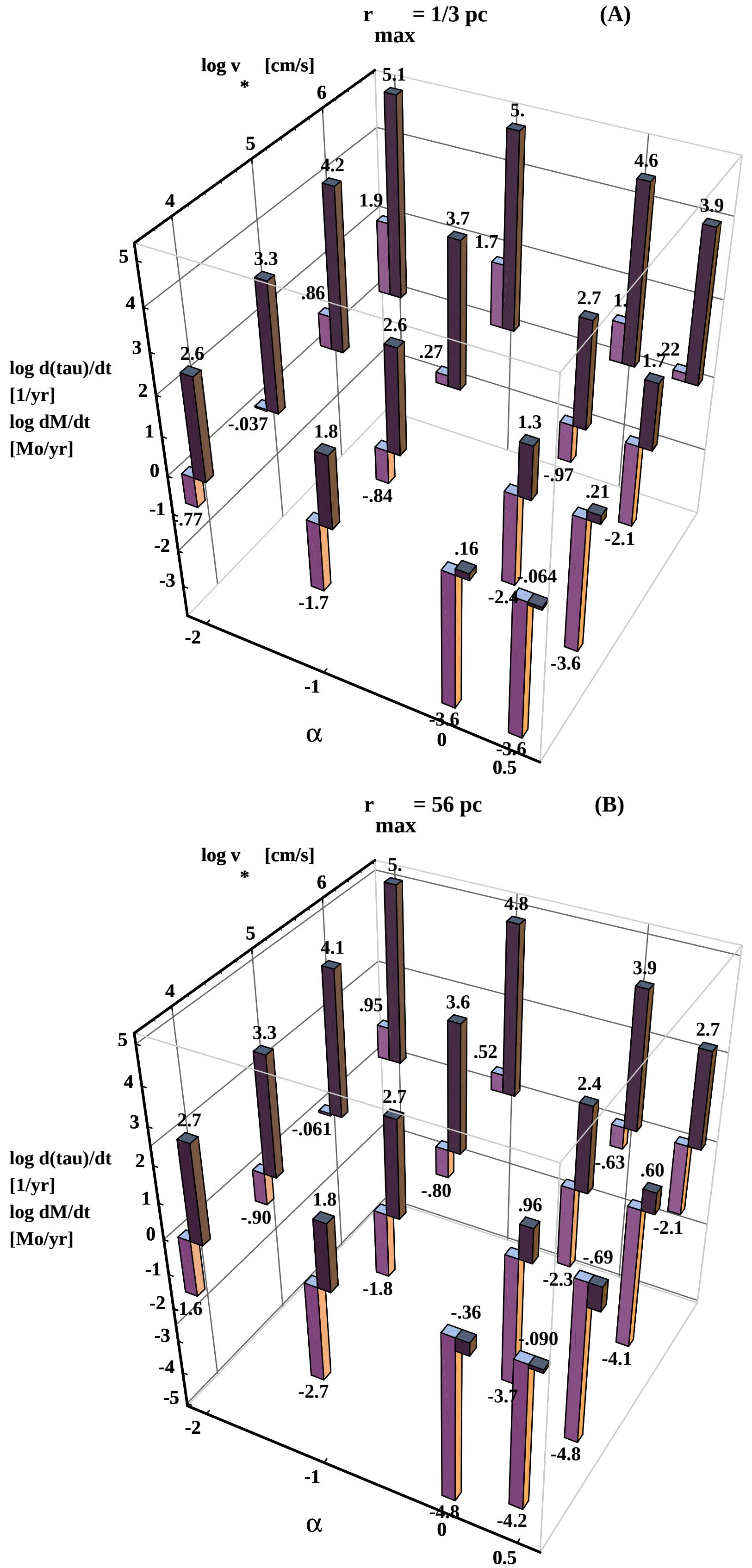

**Figure 13.** The total mass loss rate from stellar evolution, MS–MS and BS–BS collisions (dark gray columns on the right) and the buildup rate of electron scattering optical depth (light gray columns on the left). The numeric values of $\log(\dot{M})$ and $\log(\dot{\tau})$ are given above (if positive) or below (if negative) the respective columns. A) $r_{\rm out} = 1/3\,{\rm pc}$. B) $r_{\rm out} = 56\,{\rm pc}$.

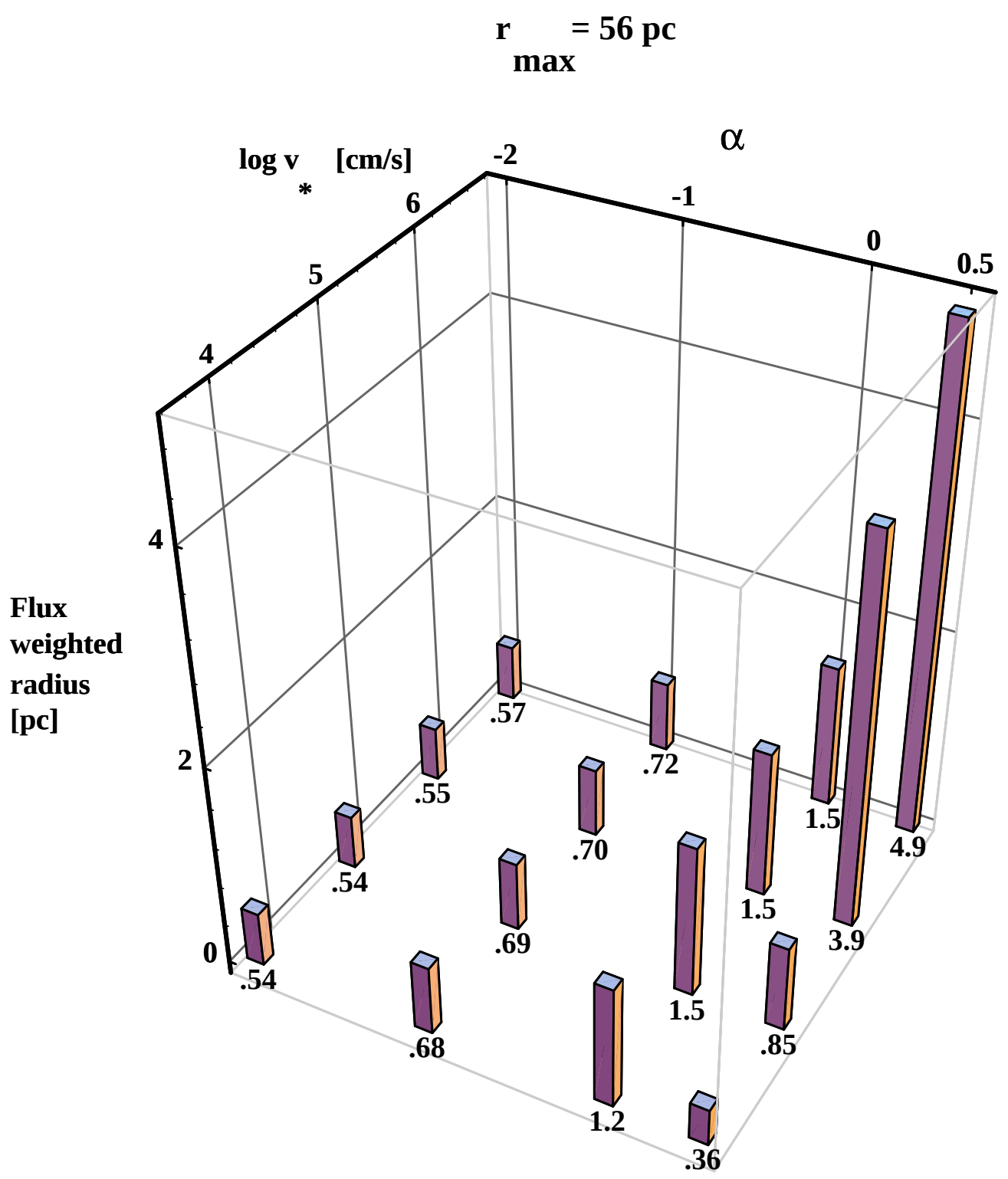

**Figure 14.** Flux weighted radius of the H$\beta$ line. $r_{\rm out} = 56\,$pc.

## 6.2 Main results

The response of the emission lines to changes of the wind structure was shown to be mostly explicable in terms of the overall properties of the wind, that is the boundary density and the relative sizes of the wind and the star. This inspires confidence in our assumption that the results, which are based on wind structure models that lack hydrodynamical self consistency, are a fair approximation of the line emission from realistic bloated stars. The dependence of the emergent line spectrum on the conditions at the boundary of the wind suggests that the central issue is to understand the mechanisms that define the outer limit of the line emitting region rather than the detailed structure of the wind.

Broad forbidden lines are not observed in AGN. The emission of forbidden lines from BSs, and in particular lines of high collisional de-excitation density such as [Fe XI] $\lambda$7892, severely restrict possible wind models. Our photoionization calculations show that the gas at the comptonized boundary has a high ionization parameter ($U \sim 3$) and emits strong broad forbidden lines. Collisional de-excitation becomes efficient only very near the black hole, well within the BLR radius. This result remains unchanged when a softer continuum is used but may be sensitive to the comptonization temperature. By artificially lowering the temperature from $\sim 6\times10^4$K to $3\times10^4$ the forbidden lines can be suppressed up to $\sim$ 1pc. Barring that, only mass limited wind models whose boundary density is higher than the comptonization density avoid excessive broad forbidden line emission.

The most conspicuous discrepancy between the calculated and observed line spectrum is the deficiency of the Mg II $\lambda$2798/L$\alpha$ and N V $\lambda$1240/L$\alpha$ ratios, regardless of the wind model or BLR size cutoff. (The Mg II $\lambda$2798/L$\alpha$ emission of dense wind models does rise up to the marginally acceptable value of $\sim$ 0.1, but this emission originates too far from the black hole). The Mg II $\lambda$2798 deficiency can be corrected by a combination of a lower optical depth and / or a lower ionization parameter. Our photoionization calculations assume that the transverse optical depth is infinite (slab geometry), which actually over-estimates the true optical depth of a spherical geometry.

The N V $\lambda$1240 deficiency in our models is similar to a general trend exhibited by cloud models. Hamann & Ferland (1993) propose that the source of this problem lies in the erroneous assumption of a solar chemical composition and suggest a composition of higher metalicity for high luminosity AGN. This is likely to improve the fit but was not included in our modeling. The N V $\lambda$1240 line is strong in high density gas (note the steep increase at small radii in fig. 5). Its deficiency may be related to our choice of $r_{\rm in}$, which is perhaps too large and artificially excludes the innermost line emitting region of the BLR.

We find, as expected, that the efficiency of the BSs in converting the irradiating continuum into line emission decreases with $r_{\rm bh}$ as an increasingly larger fraction of the wind becomes optically thin and that this effect becomes more pronounced when the wind's density gradient is steeper. The efficiency (defined as the ratio between the ionizing and the reprocessed luminosities) becomes a sensitive function of the mean wind density at high wind densities and reaches values of $C_{\rm F} \gtrsim 1$ for the densest wind model considered here. We find that the wind structure, parameterized by the wind base velocity and velocity gradient, strongly affects $EW_\star$(L$\alpha$) by controlling both the wind density and its size. The slow decelerating winds have favorable conditions for high L$\alpha$ emission because the relatively flat density gradient results in very large comptonization radii and the high density increases the efficiency of the continuum reprocessing.

The number of BS required for the BLR is directly related to the L$\alpha$ emission from the single BS, $EW_\star$(L$\alpha$). It is therefore also a very sensitive function of the wind density structure and changes by more than 4 orders of magnitude across the parameter space. We find that if we assume $r_{\rm out} = 1/3\,$pc, even the relaxed constraints on the fraction of BSs, $f_{\rm BS} \lesssim 1$, on the total mass loss from the BSs, $\dot{M} \lesssim 1 M_\odot/\,$yr and on the growth rate of the Compton optical depth, $\dot{\tau}_{\rm es} \lesssim 0.01\,{\rm yr}^{-1}$ limit the acceptable wind models to the dense slow models ($\alpha \ge 0$, $v_\star < 0.5\,$km/s or $\alpha > 0$, $v_\star < 1\,$km/s). Begelman & Sikura (1991) estimate the supergiant (SG) collisional mass loss rates for $r_{\rm BLR} = 0.1$pc and 1/3pc at $\sim$ 8500 and $\sim 50 M_\odot/\,$yr respectively and conclude that SGs cannot play the role of BSs. Our results show that these values underestimate in fact the total mass loss rate from SG-like BSs ($\alpha = 0$, $v_\star = 10\,$km/s) by up to 2 or-ders of magnitude. We therefore also conclude that SGs are ruled out by excessive mass loss which is mostly collisional.

Generally, the number of BS in the successful wind models is much smaller than the number of clouds in cloud models. The densest wind model ($\alpha = 0.5$, $v_\star = 50\,\mathrm{m/s}$) requires $\lesssim 5{\times}10^4$ BSs in the inner 1/3 pc. It is interesting to note that Atwood, Baldwin & Carswell (1982) obtained from the structure of the H$\alpha$ and H$\beta$ line profiles of Mkn 509 a *lower* limit of $5{\times}10^4$ on the number of individual line emitting objects.

The stellar core considered in this work extends well beyond the BLR size that is inferred from line reverberation studies. Wind models that produce significant line emission at $r_{\mathrm{BLR}} < r_{\mathrm{bh}} < r_{\mathrm{NLR}}$ require an additional mechanism to suppress or obscure this emission. While the need for an additional mechanism beyond the scope of the BS scenario is not necessarily an objection to this model, we note that the dense winds do not encounter this problem since the mass cutoff on the size of the wind introduces a natural cutoff to the size of the BLR by decreasing $C_{\mathrm{F}}(r_{\mathrm{bh}})$. We find that for such wind models this natural cutoff is consistent with $r_{\mathrm{BLR}}$.

Different criteria point to the slow dense models as the best candidates for realizing the bloated stars scenario: They produce the best fit to the observed emission line spectrum, they require a small number of BSs that can be supported by the stellar population and avoid rapid excessive mass loss and they need no additional mechanism for limiting the size of the BLR. The mean wind density ($10^8 \lesssim N \lesssim 10^{12}\,\mathrm{cm/s}$) and its small density gradient resemble those of pressure confined clouds and therefore the good fit to the observed spectrum is, with hindsight, not surprising. Neither is the fact the densest winds are found to be the most efficient in converting the ionizing continuum into lines. The similarity between the successful wind models and the clouds demonstrates the extent to which the observed data and the atomic physics constrain any possible BLR models. The most significant result in this feasibility study is that only a small number of BSs with dense winds are required for the BLR line emission.

The dense wind models do not resemble solar neighborhood SGs and to date no such objects were observed in any astronomical environment. However, at our current level of knowledge, the assumption that the AGN stellar population resembles the solar neighborhood population is no more likely than the opposite assumption that hitherto unobserved bloated stars exist very near the black hole.

### 6.3 limitations, uncertainties and future work

Of the many assumptions that are required to construct a physical realization of the bloated stars scenario, we list below those that we regard as potentially problematic or those that were not studied in detail in this work.

This work is limited to a specific choice of the evolutionary epoch of the core, and with it the black hole's mass and the AGN luminosity (§ 2.1). In addition, the MCD stellar core model we employ is based on a stellar population that does not include BSs. By applying their results to the BSs we are implicitly assuming that the dynamical evolution of the stellar core proceeds independently of their existence. While this may be true for those wind models that require only a small fraction of BSs in the stellar population, the existence of a significant fraction of BSs must affect the dynamics and evolution of the core.

We have ignored the effects of the irradiating continuum on the wind density structure and the assumed that the wind properties are independent of $r_{\mathrm{bh}}$. The uncertainties in the details of the processes that determine the wind boundary: the choice of the comptonization temperature $T_C$, the amount of mass in the wind, the tidal disruption radius and the effect of collisions on the wind, are problematic. The idea that mass-limited winds have a sharp, well defined boundary with a density above the comptonization density plays an important role in the collisional suppression of the unobserved broad high excitation forbidden lines. It is however not clear that such a sharp boundary is physically realistic and a more gradual transition from the wind density to the HIM density may result in a higher forbidden line emission. The actual photoionization calculations also involve several assumptions, of which the most problematic is the infinite slab approximation. Our model is minimalistic in that it assumes that the BLR is composed of a single component—the bloated stars. We do not consider for example the possibility of line emission from dusty clouds or from the accretion disk. In addition this model does not address the issue of different dynamics for different lines (e.g. the observed redshift differences between various lines).

There are two important tests that should be applied to the BS scenario in its present level of detail: modeling the line profiles and simulating line reverberation mapping of the BLR with BSs. The question whether the BS model can reproduce the observed profiles and reverberation radii is crucial for establishing its viability as an alternative to the BLR cloud models. This work is currently under way. In addition, steps towards obtaining a hydrodynamically self-consistent wind model are required, but achieving physical self-consistency in this issue is expected to involve considerable difficulties.

## 7 SUMMARY

We employed a detailed photoionization code and utilized a detailed model of the stellar core to study the possibility that the BLR clouds are in fact the envelopes of bloated stars. The main issues we investigated in this work were

(i) How is the emission line spectrum affected by the structure of the BS?
(ii) Can bloated stars reproduce the BLR emission line spectrum?
(iii) How many bloated stars are required and can the stellar population supply them?
(iv) Can the BSs avoid rapid collisional destruction and excessive mass injection to the interstellar medium?

Our results show that

(i) The line emission spectrum is mainly determined by the conditions at the outer gas layers. This makes the physical mechanisms that determine the boundary density of the wind a central issue.
(ii) Dense winds ($10^8 \lesssim N \lesssim 10^{12}\,\mathrm{cm}^{-3}$) can reproduce the observed BLR emission line spectrum

with a success comparable to that of cloud models. The main problems are the under-production of Mg II λ2798 and N V λ1240. Strong forbidden lines, especially [Fe XI] λ7892, exclude the possibility of lower density winds.

(iii) The required number of BSs is a very sensitive function of the density structure of the wind. For dense winds it is small enough to be supported by the stellar population. BSs with such winds can also avoid rapid collisional destruction and the resultant high optical depth of the ISM.

**Acknowledgements**

This work has benefited from discussions with D. Kazanas, J. Taylor, D. Lin, G. Ferland and A. Pohl. We would like to thank B. Murphy for providing us with unpublished numeric results and helpful comments. We acknowledge financial support by the US-Israel Binational Science Foundation grant no. 89-00179.